\documentclass[12pt]{article}
\usepackage[letterpaper,textwidth=6in,top=25mm,bottom=25mm]{geometry}
\usepackage[T1]{fontenc}
\usepackage[utf8]{inputenc}
\usepackage{lmodern}
\usepackage{microtype}
\usepackage{amsmath,amssymb,mathtools,bm}
\usepackage{booktabs,array}
\usepackage{xcolor}
\usepackage[numbers,sort&compress]{natbib}
\usepackage[colorlinks=true,linkcolor=blue,citecolor=blue,urlcolor=blue]{hyperref}
\allowdisplaybreaks
\numberwithin{equation}{section}

\newcommand{\AdS}{\mathrm{AdS}}
\newcommand{\vol}{\mathrm{vol}}
\newcommand{\dd}{\mathrm{d}}
\newcommand{\cL}{c_{\mathrm L}}
\newcommand{\cR}{c_{\mathrm R}}

\newcommand{\ellp}{\ell_{\mathrm p}}
\newcommand{\fluxscale}{\mathcal A}

\title{Beyond the polynomial ansatz:\\
regular $\AdS_3\times M_8$ solutions of M-theory}
\author{\begin{tabular}{c}
\normalsize Carolina Matt\'e Gregory\\[1em]
\small Universidade de Bras\'ilia, Instituto de F\'isica,\\
\small 70910-900, Bras\'ilia, DF, Brasil\\[0.4em]
\small International Center of Physics, C.P. 04667, Bras\'ilia, DF, Brasil
\end{tabular}}
\date{}

\begin{document}
\maketitle
\begin{abstract}
We study regular compact solutions of the cohomogeneity-one
$\AdS_3\times M_8$ system of Gauntlett, Mac Conamhna, Mateos and Waldram,
where $M_8$ is an $S^2$ bundle over three constant-curvature Riemann surfaces.
We find that an endpoint argument rules out every non-flat curvature-sign sector without
a positive-curvature factor, while an analytic deformation gives regular
families with four radial shape parameters in all remaining sectors,
with the base geometry held fixed. The known polynomial solutions
form a proper analytic locus in these families. Non-polynomiality is
therefore generic in parameter space, although the deviation from an
endpoint-matching polynomial can be very small. Apart from the completely
flat exception, the curvature-sign criterion is necessary and sufficient
within this ansatz. Four magnetic flux ratios give local coordinates near
the reference solution and its small-curvature deformations, and a rational
inverse reconstructs candidate endpoint data directly from finite flux
ratios. In the fully positive sector we prove that one explicit
four-dimensional finite-ratio neighbourhood lies inside the physical flux
image. The magnetic periods and leading holographic central charge reduce
to endpoint data, and, once the common Brown--Henneaux normalization is fixed,
the nontrivial flux dependence of the equivariant localization formula
agrees with the endpoint result as an exact algebraic identity. We also
quantify the non-polynomial corrections and analyse finite degenerations
under explicit convergence hypotheses.
\end{abstract}
\clearpage
\begingroup
\footnotesize
\setcounter{tocdepth}{2}
\tableofcontents
\endgroup
\clearpage

\section{Introduction}\label{introduction}

Supersymmetric $\AdS_3$ backgrounds relate the geometry of wrapped branes
to two-dimensional conformal field theory
\cite{MaldacenaNunez:2000,Gauntlett:2006m5,BeniniBobev:2012,BeniniBobev:2013}.
The local supersymmetry conditions are known in considerable generality
\cite{OColgainEtAl:2010}, while regular compact solutions require a
separate global construction. Figueras, Mac Conamhna and \'O Colg\'ain
\cite{FiguerasEtAl:2007} formulate the global spinorial geometry of a
broad class of $\AdS_3$ backgrounds, and Ashmore \cite{Ashmore:2023}
gives necessary and sufficient local supersymmetry conditions for general
$\mathcal N=(2,0)$ $\AdS_3\times M_8$ solutions in terms of an
$SU(2)$ structure. Those
classifications do not solve the global two-pole boundary-value problem
studied here. We instead work in the cohomogeneity-one sector of Gauntlett,
Mac Conamhna, Mateos and Waldram (GMMW) \cite{Gauntlett:2006ux}, where
$M_8$ is an $S^2$ bundle over
$B_6=C_1\times C_2\times C_3$. GMMW constructed explicit regular
families in this sector. Our question is which endpoint-compatible data
admit regular compact fillings without imposing their polynomial ansatz.
Each $C_i$ is a compact constant-curvature Riemann surface, and the field
equations reduce to a nonlinear second-order equation for a master function
$H(r)$.

The GMMW sector has a precise wrapped-brane origin, although its internal
geometry is more general than the standard wrapped-M5 near-horizon ansatz.
In section 2 and appendix A.2 of \cite{Gauntlett:2006ux}, the relevant
supersymmetry conditions are derived from the Killing-spinor projections of
M5-branes wrapping K\"ahler four-cycles in a Calabi--Yau fourfold.  The
familiar wrapped-M5 geometry contains an $S^4$ in the transverse directions.
GMMW motivate their broader construction by generalizing this transverse
geometry to an $S^2$ bundle over a two-dimensional K\"ahler factor, and then
allow the six-dimensional base to be a product
of three two-dimensional K\"ahler--Einstein spaces, a $KE_6$ space, or a product $KE_4\times KE_2$.  The same identification
is made explicitly in section 4.3 of \cite{Ashmore:2023}.  The resulting
G-structure class may also carry an electric component of $G_4$, so its
wrapped-M5 origin should not be read as an assertion that every solution is
sourced by M5-branes alone.  With this qualification, the compact GMMW
solutions are holographic duals of chiral two-dimensional SCFTs with
$\mathcal N=(0,2)$ supersymmetry (with the opposite boundary orientation
used here, this is $\mathcal N=(2,0)$).  We will not attempt to identify a
microscopic SCFT for each non-polynomial geometry.  Rather, the magnetic
flux lattice, the leading central charge and the gravitational anomaly are
the robust holographic data that can be followed throughout the families
constructed in what follows.

The polynomial solutions of GMMW provide explicit regular backgrounds, presented as constructions rather than as a classification. Their reported compact classes
cover three of the six curvature-sign sectors reached below. In particular,
the $(+,+,+)$, $(+,0,0)$ and $(+,-,0)$ sectors are absent from their list
of regular examples, as detailed after \eqref{np-analytic-all-signs}.
Here we solve the underlying two-endpoint problem without a polynomial
ansatz. At fixed base metrics, their volume factors and the bundle connection,
quotienting radial translation and homogeneous scaling leaves four radial
shape parameters. This count does not include Teichm\"uller parameters,
variations of flat connection data, or flat $C$-field choices. The count already suggests what to
expect from non-polynomiality. Once a four-dimensional regular family has
been constructed, the polynomial equation cuts out a proper analytic locus,
locally a hypersurface, where its defining obstruction has nonzero differential,
so a generic nearby regular solution is non-polynomial. Here and below, genericity refers to codimension in the
regular parameter space, not to the distance from the polynomial locus. A
non-polynomial solution may sit extremely close to an endpoint-matching
polynomial profile.

We first construct these four parameters analytically near an exact
$S^2\times T^2\times T^2$ solution. The linearized boundary-value operator
is coercive, and the endpoint conditions remove the resonance that could
obstruct smoothness at the poles. A separate polynomial obstruction shows
that the exceptional polynomial locus is genuinely proper.
Small-curvature deformations enter every sign sector containing a positive factor, while an endpoint
argument excludes all other non-flat sectors. The curvature-sign criterion
is therefore sharp within this ansatz.

We next turn to the magnetic fluxes. The four-ratio Jacobian is exactly
nonsingular at the reference point and remains so in a sufficiently small
analytic neighbourhood. Rational flux shapes there give infinitely many
shifted-flux-quantized backgrounds. At fixed genus, however, the
small-curvature construction runs to large physical flux aspect ratios, and
curvature normalization does not change those ratios. We therefore make no
claim from this perturbative construction about generic finite-ratio
fluxes. Instead, a rational inverse reconstructs candidate endpoint data,
and, in the fully positive sector, we prove directly that one explicit
four-dimensional finite-ratio neighbourhood lies inside the physical flux
image.

Finally, the magnetic periods and the leading Brown--Henneaux central
charge both reduce to endpoint data. We specialize the equivariant
localization formulae
\cite{BenettiGenoliniEtAl:2023PRL,BenettiGenoliniEtAl:2023M5,BenettiGenoliniEtAl:2024}
and prove algebraically that, once the common Brown--Henneaux normalization
is fixed, their nontrivial flux dependence agrees with the direct endpoint
expression. This identity does not require interior existence, and the
gravitational anomaly is likewise determined by the flux class and the
projective-bundle topology. We use the $\mathcal N=(2,0)$ boundary
orientation throughout.

One would find that \cite{CompanionLocalisation} has a different emphasis. There, we develop the general equivariant-localization framework for
$\AdS_3\times M_8$ backgrounds and use the same quantized $(+,+,0)$
non-polynomial geometry as an explicit application. Here, the focus is the
global GMMW boundary-value problem, i.e., the curvature-sign criterion, analytic
existence, the flux map, one finite-ratio image patch, and finite
degenerations. The two papers overlap only in the benchmark geometry and in
the endpoint observables used to compare the descriptions.

The paper is organized as follows. In section~\ref{unrestricted-system},
we review the GMMW ansatz, the endpoint regularity conditions, and the
magnetic fluxes. In section~\ref{analytic-classification}, we establish the
curvature-sign criterion and construct the regular analytic families. In
section~\ref{flux-space}, we reconstruct candidate endpoint data from fluxes
and establish one finite-ratio image patch in the fully positive sector. In
section~\ref{explicit-representatives}, we present explicit representatives
and quantify their departure from polynomial profiles. In
section~\ref{endpoint-observables}, we reduce the central charge and anomaly
to endpoint data and compare with equivariant localization. In
section~\ref{global-structure}, we discuss numerical continuation and finite
degeneration mechanisms. We conclude in section~\ref{discussion}. The
appendices collect the curvature normalization, endpoint analysis, the
finite-ratio existence proof, exact flux differentiation, finite-degeneration
proofs, and numerical continuation data.

\section{Geometry and endpoint data}\label{unrestricted-system}
\label{np-gmmw-ansatz}

The bosonic fields of eleven-dimensional supergravity are the metric and
the four-form field strength $G_4$. They obey the Bianchi identity
$\dd G_4=0$, the four-form equation
$\dd *_{11}G_4+\tfrac12G_4\wedge G_4=0$, and the Einstein equation.
We restrict these equations, together with the Killing-spinor conditions
for $\mathcal N=(2,0)$ supersymmetry, to the cohomogeneity-one GMMW
ansatz of \cite{Gauntlett:2006ux}. The internal space is locally an
$S^2$ fibration over $B_6=C_1\times C_2\times C_3$. The coordinate $r$
parametrizes the orbit space of the fibre $U(1)$ action, while $\psi$ is
its angular coordinate. Global compactness will require the $\psi$ circle
to collapse smoothly at two endpoints of the $r$ interval.

Supersymmetry reduces the coupled supergravity system very substantially, and once the three quadratic functions $Q_i$ below are specified, all metric
and flux functions are reconstructed explicitly from a single radial
function $H(r)$. The remaining differential condition is one nonlinear
second-order equation for $H$, supplemented by the endpoint and positivity
conditions imposed below. Let $\operatorname{Ric}(C_i)=k_i g_{C_i}$ and
let $J_i$ be the corresponding K\"ahler forms. In the conventions of
\cite{Gauntlett:2006ux}, the dimensionless metric and four-form are
\begin{align}
 ds_{11}^2&=\omega^2\left[ds^2(\AdS_3)+\sum_i h_i ds^2(C_i)
                  +f_3\,dr^2+f_4(D\psi)^2\right],
 \label{np-gmmw-metric}\\*
 G_4&=\sum_{\rm cyc}g_iJ_j\wedge J_\ell
       +\sum_i g_{i+3}J_i\wedge dr\wedge D\psi
       +g_7\,dr\wedge\vol_{\AdS_3},
 \label{np-gmmw-fourform}
\end{align}
where $\{i,j,\ell\}=\{1,2,3\}$, all scalar functions depend on $r$, and
\begin{equation}
 D\psi=\dd\psi+\mathcal P,\qquad
 \dd\mathcal P=\sum_i k_iJ_i,\qquad \psi\sim\psi+2\pi.
 \label{np-gmmw-connection}
\end{equation}
Introduce the reduced polynomials and reconstruction functions
\begin{align}
 K_i&=k_jk_\ell,\qquad Q_i=K_i r^2-4c_i r-4d_i,
 &P&=Q_1Q_2Q_3,\label{np-general-Q}\\*
 R&=-\sum_i k_iQ_i,\qquad D=2H''+R,
 &\varphi&=\frac{4H'}D,\label{np-general-D}\\*
 h_i&=-\frac{D}{4Q_i},\qquad f_3=-\frac{D}{16H},
 &f_4&=-\frac{HD}{P},\qquad\omega^6=\frac{4P}{D^2}.
 \label{np-gmmw-reconstruction}
\end{align}
The flux coefficients are
\begin{align}
 g_i&=-\frac12[\varphi(k_jh_\ell+k_\ell h_j)+K_i r]+c_i,
 \qquad -4g_{i+3}=(2\varphi h_i)'+k_i,
 \label{np-gmmw-magnetic-functions}\\*
 g_7&=\frac{f_3}{2h_1h_2h_3}\left[
 (\omega^6+\varphi^2)\sum_{\rm cyc} k_i h_jh_\ell
 +\varphi r\sum_{\rm cyc}K_i h_i
 -2\varphi\sum_i c_i h_i\right].\label{np-gmmw-electric-function}
\end{align}
The remaining supersymmetry and field-equation content is
\begin{equation}
 \mathcal E_H[H]\equiv-(H')^2+H(2H''+R)+\frac14P=0.
 \label{np-general-master}
\end{equation}
We shall refer to \eqref{np-general-master} as the \textit{master equation}. Once a
solution $H$ is known, equations \eqref{np-gmmw-reconstruction}--
\eqref{np-gmmw-electric-function} reconstruct the local supergravity
background, while global regularity becomes a boundary-value problem for
$H$ and the reduced parameters.

\subsection{Regularity and radial gauge}\label{nonpoly-endpoint-formulation}
Regularity is imposed directly on solutions of the master equation
\eqref{np-general-master}. The fibre closes at two simple zeros
$r_-<r_+$ of $H$. Regularity requires $H>0$ in the open interval, $Q_i>0$
and $D<0$ on its closure, and smooth endpoint germs. Evaluating
\eqref{np-general-master} and its derivative at a pole, we find 
\begin{equation}
 H_\pm'^2=\frac14P_\pm,\qquad
 H'_\pm R_\pm+\frac14P'_\pm=0,\qquad
 H'_->0>H'_+.
 \label{np-general-endpoint-two}
\end{equation}
The first equation in \eqref{np-general-endpoint-two} is precisely the
absence of a conical singularity when the fibre coordinate has period
$2\pi$. For example, at the left endpoint, setting
$\rho^2=-D_-(r-r_-)/(4H'_-)$ gives
$f_3dr^2+f_4(D\psi)^2=d\rho^2+\rho^2(D\psi)^2+O(\rho^4)$.
Thus, the $(r,\psi)$ plane closes as an ordinary smooth polar plane rather
than as a cone.

The four-form is regular at the same pole. The coefficients $g_i$ are
finite, while
$2\varphi h_i=-2H'/Q_i$ is analytic, so the coefficients $g_{i+3}$ are
finite. Moreover, $dr=O(\rho\,d\rho)$, and therefore
$dr\wedge D\psi$ is a smooth multiple of the polar area form
$\rho\,d\rho\wedge D\psi$. For the electric term, we denote the square
bracket in \eqref{np-gmmw-electric-function} by $\mathcal B$. At a pole,
$H'^2=P/4$ gives $\omega^6=\varphi^2$, and a direct substitution yields
\begin{equation}
 \mathcal B_e
 =-\frac{R_e}{2}
 -\frac{H'_e}{2}\frac{P'_e}{P_e}=0,
 \label{np-endpoint-G4-cancellation}
\end{equation}
as $H'_eR_e+P'_e/4=0$. Hence, $\mathcal B=O(r-r_e)$ cancels the
simple pole of $f_3$, so $g_7$ is finite. The same calculation applies at
the right endpoint. Thus, both the metric and $G_4$ extend smoothly across
the two poles. The endpoint positions, then, satisfy $P'^2-4PR^2=0$, with the signs in
\eqref{np-general-endpoint-two} selecting the physical roots.

Radial translation and positive homogeneous scaling act by
\begin{align}
 \widetilde r&=r+a,&
 \widetilde c_i&=c_i+\tfrac12K_i a,&
 \widetilde d_i&=d_i-c_i a-\tfrac14K_i a^2,
 \label{np-finite-shift}\\*
 \widetilde r&=\lambda r,&
 \widetilde c_i&=\lambda c_i,&
 \widetilde d_i&=\lambda^2d_i,\qquad
 \widetilde H(\widetilde r)=\lambda^4H(r).
 \label{np-finite-scale}
\end{align}
Translation means $\widetilde H(\widetilde r)=H(\widetilde r-a)$.
We use these freedoms to set $(r_-,r_+)=(0,1)$ when studying shapes, as they leave the flux ratios below invariant. Curvature normalization is a
different equivalence, as proved in appendix~\ref{curvature-normalization}.

\subsection{Magnetic fluxes}\label{np-flux-coordinates}
The local fibre periods and pole coefficients are
\begin{equation}
 I_i=\frac12\left[\frac{H'}{Q_i}\right]_-^+
          -\frac{k_i}{4}(r_+-r_-),\qquad
 G_i^\pm=-\frac14Q_i'(r_\pm)
       +\frac12\left(k_j\frac{H'_\pm}{Q_\ell^\pm}
                     +k_\ell\frac{H'_\pm}{Q_j^\pm}\right).
 \label{np-general-fibre-periods}
\end{equation}
We restore the physical fields by
$ds^2_{11,\rm phys}=L_{\!A}^2ds^2_{11}$ and
$G_{4,\rm phys}=L_{\!A}^3G_4$, and define
\begin{equation}
 \fluxscale=\frac{L_{\!A}^3}{\pi\ellp^3},\qquad
 \nu_i=\frac1{4\pi}\int_{C_i}J_i,\qquad
 N_i=\fluxscale\nu_i I_i,\qquad M_i=2\fluxscale\nu_j\nu_\ell G_i^-.
 \label{np-local-to-integral-periods}
\end{equation}
The unshifted vector is always ordered as
$\Gamma=(N_1,N_2,N_3;M_1,M_2,M_3)$, with $N_i$ integrated over the
fibre above $C_i$ and $M_i$ over $C_j\times C_\ell$ in the south section.
For nonzero curvature, the Gauss--Bonnet theorem fixes the normalized area through
\begin{equation}
 \frac{1}{4\pi}\int_{C_i} k_i J_i=1-g_i,
 \label{gauss-bonnet}
\end{equation}
and hence $k_i\nu_i=n_i=1-g_i$. Therefore, a torus has a freely chosen positive volume.

On the chart $I_1I_3\neq0$, write
\begin{equation}
 (x,y,u,v,w)=\frac1{I_1}(I_2,I_3,G_1^-,G_2^-,G_3^-).
 \label{np-fluxmap-ratios}
\end{equation}
The endpoint equations imply
\begin{equation}
 \sum_i I_iG_i^-+\sum_{\rm cyc}k_i I_jI_\ell=0,
 \qquad u+xv+yw+k_1xy+k_2y+k_3x=0.
 \label{np-general-flux-compatibility}
\end{equation}
Thus, $(x,y,u,v)$ are four candidate shape coordinates, and, at fixed
$(k_i,\nu_i)$, physical flux ratios differ by fixed volume factors.
The local independence on regular solutions will be proved in
section~\ref{np-analytic-flux}.

\section{Curvature signs and analytic existence}\label{analytic-classification}

Within this ansatz, a non-flat curvature-sign sector admits regular compact
solutions if and only if it contains a positive factor. Moreover, every
such sector contains open families of non-polynomial solutions.
We first prove the endpoint no-go result, then construct the allowed families.

\subsection{A curvature no-go result}\label{nonpoly-nogo}

In searching for new non-polynomial solutions to the GMMW ansatz, numerical computations repeatedly failed
in the sectors in which every curvature was non-positive. Looking for the reason why that happened, we found that the endpoint
equations, in fact, completely 
exclude these sectors.

\paragraph{No-go statement.}
Normalize every nonzero curvature to $k_i=\pm1$.  If
\begin{equation}
 k_i\leq0\qquad\hbox{for all }i,
 \label{np-nogo-nonpositive}
\end{equation}
and at least one $k_i$ is strictly negative, then there is no smooth
compact two-endpoint solution with positive internal metric. Here each
$C_i$ is a compact constant-curvature Riemann surface and the $S^2$
fibre closes smoothly at two ordinary poles with $\psi\sim\psi+2\pi$.
It is worth noticing that this statement does not cover orbifold pole closures or more general
internal geometries. Equivalently, apart from the completely flat branch
discussed below, every regular compact solution in this ansatz contains
at least one positively curved two-dimensional factor.

The proof uses only the endpoint equations and positivity of the metric.
This is worth emphasizing, because it is the converse of the existence construction
below and turns the sign classification into an ``if and only if'' result.

For a regular solution, $H>0$ in the open interval.  From
\eqref{np-gmmw-reconstruction}, positivity of the radial and base
coefficients implies
\begin{equation}
 D<0,\qquad Q_i>0
 \label{np-nogo-basic-signs}
\end{equation}
on the closed interval.  Radial translation and positive homogeneous
scaling allow us to put
\begin{equation}
 r_-=0,\qquad r_+=1
 \label{np-nogo-unit-interval}
\end{equation}
without changing any sign condition.  Define
\begin{equation}
 x_i=\sqrt{Q_i(0)},\qquad y_i=\sqrt{Q_i(1)},\qquad
 X=x_1x_2x_3,\qquad Y=y_1y_2y_3.
 \label{np-nogo-xy}
\end{equation}
Then $x_i,y_i>0$, and the endpoint slopes are
$H'(0)=X/2$, $H'(1)=-Y/2$.  With $W=\sqrt P$, the differentiated
endpoint condition in \eqref{np-general-endpoint-two} becomes
\begin{equation}
 W'(0)=-R(0),\qquad W'(1)=R(1).
 \label{np-nogo-W-endpoint}
\end{equation}

Because each $Q_i$ is quadratic, its values in the interval are determined
by its endpoint values,
\begin{equation}
 Q_i(r)=[(1-r)x_i-r y_i]^2
 +r(1-r)\bigl[(x_i+y_i)^2-K_i\bigr].
 \label{np-boundary-Q-square}
\end{equation}
When $K_i=1$, positivity therefore forces
\begin{equation}
 x_i+y_i>1.
 \label{np-nogo-triangle}
\end{equation}
This elementary inequality is the key input in the all-negative and
two-negative cases.

First take
\begin{equation}
 (k_1,k_2,k_3)=(-1,-1,-1).
 \label{np-nogo-mmm}
\end{equation}
Here every $K_i=1$ and
\begin{equation}
 Q_i(r)=(1-r)x_i^2+r y_i^2-r(1-r).
 \label{np-nogo-quadratic}
\end{equation}
Set
\begin{equation}
 a_i=\frac{x_i^2+1-y_i^2}{2x_i},\qquad
 b_i=\frac{y_i^2+1-x_i^2}{2y_i}.
 \label{np-nogo-ab-quadratic}
\end{equation}
Since $R=Q_1+Q_2+Q_3$, the two equations
\eqref{np-nogo-W-endpoint} are equivalent to
\begin{equation}
 \sum_i a_i x_jx_\ell=\sum_i x_i^2,\qquad
 \sum_i b_i y_jy_\ell=\sum_i y_i^2,
 \qquad \{i,j,\ell\}=\{1,2,3\}.
 \label{np-nogo-endpoint-mmm}
\end{equation}
For each $i$ one has the exact identity
\begin{equation}
 \frac1{x_iy_i}
 -\left(\frac{a_i}{x_i}+\frac{b_i}{y_i}\right)
 =
 \frac{(x_i-y_i)^2\bigl[(x_i+y_i)^2-1\bigr]}
 {2x_i^2y_i^2}\geq0,
 \label{np-nogo-key-quadratic}
\end{equation}
where the sign follows from \eqref{np-nogo-triangle}.  The same inequality
also gives the strict estimate
\begin{equation}
 \frac1{x_iy_i}<\frac1{x_i}+\frac1{y_i}.
 \label{np-nogo-strict}
\end{equation}
Finally,
\begin{equation}
 \sum_i\frac1{x_i}
 =\frac{x_1x_2+x_1x_3+x_2x_3}{X}
 \leq\frac{x_1^2+x_2^2+x_3^2}{X},
 \label{np-nogo-square-ineq}
\end{equation}
and similarly for the $y_i$. The difference of the two numerators is
$\frac12\sum_{i<j}(x_i-x_j)^2$.  Dividing the two equations in
\eqref{np-nogo-endpoint-mmm} by $X$ and $Y$ and adding them, the left-hand
side is bounded above by $\sum_i1/(x_iy_i)$, while the right-hand side is
bounded below by $\sum_i(1/x_i+1/y_i)$.  Equation
\eqref{np-nogo-strict} makes these inequalities incompatible.  Hence the
$(-,-,-)$ sector contains no regular compact solution.

Next consider
\begin{equation}
 (k_1,k_2,k_3)=(-1,-1,0)
 \label{np-nogo-mm0}
\end{equation}
up to permutation.  Now $Q_1,Q_2$ are linear, whereas $Q_3$ is the
quadratic \eqref{np-nogo-quadratic}, so $x_3+y_3>1$.  For a linear
factor define
\begin{equation}
 a_i=\frac{x_i^2-y_i^2}{2x_i},\qquad
 b_i=\frac{y_i^2-x_i^2}{2y_i}.
 \label{np-nogo-ab-linear}
\end{equation}
Then
\begin{equation}
 \frac{a_i}{x_i}+\frac{b_i}{y_i}
 =-\frac{(x_i-y_i)^2(x_i+y_i)^2}{2x_i^2y_i^2}\leq0.
 \label{np-nogo-linear-ineq}
\end{equation}
Since $R=Q_1+Q_2$, division and addition of the endpoint equations gives
\begin{align}
 0
 &\leq
 \frac1{x_3y_3}
 -\frac{x_1^2+x_2^2}{X}
 -\frac{y_1^2+y_2^2}{Y}\notag\\
 &\leq
 \frac{1-2(x_3+y_3)}{x_3y_3}
 <0,
 \label{np-nogo-mm0-contradiction}
\end{align}
where $x_1^2+x_2^2\geq2x_1x_2$ and the analogous inequality for
$y_1,y_2$ were used.  This contradiction excludes the $(-,-,0)$ sector.

For
\begin{equation}
 (k_1,k_2,k_3)=(-1,0,0)
 \label{np-nogo-m00}
\end{equation}
all three $Q_i$ are linear and $R=Q_1$.  Every contribution of the form
\eqref{np-nogo-linear-ineq} is non-positive.  After the two endpoint
equations are divided by $X$ and $Y$ and brought to one side, they also
contain the strictly negative terms $-x_1^2/X$ and $-y_1^2/Y$. Their sum therefore cannot vanish, and this excludes the final non-flat
all-non-positive sign pattern.  The three cases above exhaust the
forbidden sectors up to permutation.

The completely flat case is different.  When
$(k_1,k_2,k_3)=(0,0,0)$, one has $R=0$ and all $Q_i$ are linear.  The
endpoint conditions force equality in \eqref{np-nogo-linear-ineq}, hence
$x_i=y_i$ for all $i$.  The $Q_i$ are therefore positive constants and
the master equation admits
\begin{equation}
 H(r)=\frac{\sqrt P}{2}r(1-r).
 \label{np-nogo-flat-H}
\end{equation}
This is a smooth exceptional stratum rather than a limiting member of the
non-flat no-go statement.

For compact orientable constant-curvature surfaces, positive, zero and
negative curvature correspond respectively to genus $0$, $1$ and
$g\geq2$.  Thus, every non-flat regular compact background in the present
$C_1\times C_2\times C_3$ ansatz contains an $S^2$ factor.  Note that this conclusion is specific to the cohomogeneity-one system studied here, and not a no-go theorem for general $\AdS_3\times M_8$ solutions of
eleven-dimensional supergravity.

\subsection{A regular analytic family}\label{np-analytic-functional}

The no-go argument gives necessity, and to prove that the remaining sign
sectors really occur, we start from an exact regular solution in the
$(+,0,0)$ chamber and solve the full two-endpoint problem perturbatively.
There are two logically separate implicit-function steps.  The first
solves the two algebraic endpoint constraints inside the six-dimensional
space of quadratic data. The second solves the nonlinear differential
equation for the interior profile with those endpoint data held fixed.
Keeping these steps separate is useful later when fluxes are used to
reconstruct candidate endpoints, as the endpoint algebra by itself does not guarantee a regular filling.

For
\begin{equation}
 (k_1,k_2,k_3)=(1,0,0),\qquad r_-=0,\qquad r_+=1,
 \label{np-reference-curvatures}
\end{equation}
an exact reference solution is
\begin{equation}
 Q_1=\frac18,\qquad Q_2=1+r,\qquad Q_3=2-r,\qquad
 H_0=\frac14r(1-r),\qquad D_0=-\frac98.
 \label{np-reference-solution-functions}
\end{equation}
All three $Q_i$ are positive, $H_0>0$ in the open interval and the
constant value of $D_0$ is strictly negative.  Thus, every metric
coefficient in \eqref{np-gmmw-reconstruction} has a positive margin.
The endpoint slopes are $H_0'(0)=1/4$ and $H_0'(1)=-1/4$.

Write
\begin{equation}
 Q_i=A_i+B_i r
 \label{np-analytic-linear-Q}
\end{equation}
in this chamber and use
$q=(A_2,B_2,A_3,B_3)$ near
$q_0=(1,1,2,-1)$.  The two remaining coefficients $A_1,B_1$ are
determined by the endpoint equations.  Since $R=-Q_1$, these equations
may be written as
\begin{equation}
 F_-=W'(0)-A_1=0,\qquad
 F_+=W'(1)+(A_1+B_1)=0,\qquad W=\sqrt P.
 \label{np-analytic-endpoint-F}
\end{equation}
At the reference point,
\begin{equation}
 \frac{\partial(F_-,F_+)}{\partial(A_1,B_1)}\bigg|_0
 =\begin{pmatrix}-\tfrac12&2\\[1mm]\tfrac12&\tfrac52\end{pmatrix},
 \qquad \det=-\frac94.
 \label{np-analytic-endpoint-J}
\end{equation}
The finite-dimensional analytic implicit-function theorem therefore gives
unique analytic functions $A_1(q)$ and $B_1(q)$ near $q_0$. This just solves
the two pole-compatibility conditions and reduces the six coefficients of
the linear $Q_i$ to the expected four quotient shape parameters.

The remaining question is whether each nearby endpoint-compatible set of
$Q_i$ is filled by a regular solution of the master equation.  The two
prescribed zeros of $H$ are removed by setting
\begin{equation}
 s=r(1-r),\qquad
 \xi=\log\frac{r}{1-r},\qquad
 H=sY,
 \label{np-analytic-regularization}
\end{equation}
making the compact endpoints $\xi=\mp\infty$.  Direct substitution into
\eqref{np-general-master} gives
\begin{equation}
 \mathcal E(Y;q)
 =2YY_{\xi\xi}-Y_\xi^2-Y^2+sRY+\frac14P=0,
 \qquad
 Y(\mp\infty)=y_\mp=\frac{\sqrt{P(r_\mp)}}2.
 \label{np-analytic-Y-equation-general}
\end{equation}
The endpoint constants are not additional free data. Instead, they are fixed by the
regular pole slopes.  In particular, at the reference solution $Y_0=1/4$.

To place the problem on a fixed Banach space, let
$b_q=(1-r)y_-+r y_+$ be the elementary interpolation between the two
asymptotic constants, and write $Y=b_q+\eta$.  Because
$H^2(\mathbb R)\hookrightarrow W^{1,\infty}(\mathbb R)$, all quadratic
products in $\mathcal E$ define an analytic map
\begin{equation}
 (q,\eta)\longmapsto \mathcal E(b_q+\eta;q):
 \mathbb R^4\times H^2(\mathbb R)\longrightarrow L^2(\mathbb R).
 \label{np-analytic-map}
\end{equation}
The apparent constant terms at $\xi=\pm\infty$ cancel because
$y_\pm^2=P_\pm/4$, while $s$ and the remaining inhomogeneous terms decay
exponentially.  This is why the endpoint equations are imposed before the
functional implicit-function theorem.

Linearizing at $(q_0,Y_0)$ gives the one-dimensional Schr\"odinger-type
operator
\begin{equation}
 \mathcal D_{\rm lin}\eta
 =\frac12\eta_{\xi\xi}
 -\left(\frac12+\frac18s\right)\eta.
 \label{np-analytic-linear-operator}
\end{equation}
Its coercivity is immediate, i.e.,
\begin{equation}
 -\langle\eta,\mathcal D_{\rm lin}\eta\rangle
 =\frac12\|\eta_\xi\|_2^2
 +\int_{\mathbb R}\left(\frac12+\frac18s\right)\eta^2\,\dd\xi
 \geq\frac12\|\eta\|_{H^1}^2.
 \label{np-analytic-coercive}
\end{equation}
Lax--Milgram gives a unique weak solution for every $L^2$ source, and
the differential equation upgrades it to $H^2$.  Hence
$\mathcal D_{\rm lin}:H^2\to L^2$ is an isomorphism, and the analytic
implicit-function theorem now produces a unique local solution
$\eta(q)\in H^2$ of the nonlinear equation.

The Sobolev statement alone is not yet the desired geometric regularity
at $r=0,1$, because the logistic coordinate has moved the poles to
infinity.  Appendix~\ref{functional-analysis} treats the endpoint flow
directly and shows that $H=sY$ has smooth pole germs with the prescribed
slopes and depends continuously on the parameters in $C^2([0,1])$.
Consequently, \begin{equation}
 H>0,\qquad Q_i>0,\qquad D<0
 \label{np-analytic-strict-regularity}
\end{equation}
persist after shrinking the parameter neighbourhood.  This gives an open
four-parameter family of genuine regular solutions of the full
boundary-value problem.

This construction is local and does not provide a numerical radius for
the parameter neighbourhood. Away from it, endpoint data reconstructed by
section~\ref{np-flux-endpoint-inverse} require a separate interior
existence argument.

\subsection{Non-polynomiality and curvature deformations}
\label{np-analytic-nonpoly}
The four-dimensional count concerns radial shapes with the base metrics,
volume factors and bundle connection held fixed. In the gauge
$(r_-,r_+)=(0,1)$, the two pole-compatibility conditions eliminate two of
the six quadratic coefficients. The functional implicit-function theorem
then fixes the interior profile locally and uniquely for each of the four
remaining parameters $q$. Thus, the two integration constants of the
second-order master equation do not supply additional local moduli.
Polynomiality imposes a further analytic condition, namely, its zero set has
analytic codimension at least one when the condition is nontrivial, and is
a smooth hypersurface wherever its differential is nonzero. We now show
that the condition is nontrivial on the constructed family.

A polynomial solution in this sector has degree at most three, since a
term $h_n r^n$ with $n\geq4$ produces the uncancelled contribution
$n(n-2)h_n^2r^{2n-2}$ in $2HH''-(H')^2$.
The unique cubic with the prescribed zeros and slopes is
\begin{align}
 H_3=s(\alpha+\beta r),\qquad
 &\alpha=\frac{W_-}{2},\qquad\beta=\frac{W_+-W_-}{2},\notag\\*
 \mathcal E_H[H_3]=s^2\mathcal C,\qquad
 &
 \mathcal C=\beta(B_1+3\beta).
 \label{np-analytic-poly-obstruction}
\end{align}
Along $q=(1,1+\epsilon,2,-1)$, solving the exact endpoint equations gives
\begin{align}
 A_1&=\frac18+\frac14\epsilon+O(\epsilon^2),&
 B_1&=-\frac1{16}\epsilon+O(\epsilon^2),\notag\\*
 \beta&=\frac1{144}\epsilon^3+O(\epsilon^4),&
 \mathcal C&=-\frac1{2304}\epsilon^4+O(\epsilon^5).
 \label{np-analytic-deformation}
\end{align}
Thus, the polynomial obstruction is not identically zero. Its analytic
zero set has empty interior, and its complement gives regular
non-polynomial solutions. In this sense, the non-polynomiality claim
is a genericity statement inside an already-established regular family,
not evidence that the polynomial constructions of GMMW were intended to be
exhaustive. The inverse estimate near the reference point
also gives $\|H-H_3\|_\infty=O(|\mathcal C|)$, explaining the
fourth-order smallness along this particular deformation, for which
$d\beta(q_0)=0$.

Now keep $k_1=1$ and allow small $k_2,k_3$, writing
\begin{equation}
 Q_i=K_i r^2+B_i r+A_i,\qquad
 F_{\rm L}=W'(0)+R(0)=0,\qquad F_{\rm R}=W'(1)-R(1)=0.
 \label{np-analytic-general-endpoint-map}
\end{equation}
Both implicit-function arguments retain the same invertible derivatives
at the reference point and, therefore, give a joint analytic family in
$(q,k_2,k_3)$. For general curvatures, the degree comparison above excludes
all polynomial degrees greater than four. Choose a small nonzero
$\epsilon_*$ in \eqref{np-analytic-deformation}. Its solution has positive
distance from the closed subspace of quartics in $C^2([0,1])$.
Continuity then preserves non-polynomiality and regularity on a product
neighbourhood of this solution and $(k_2,k_3)=(0,0)$.

Consequently, there are $\delta>0$ and an open set of shape parameters
for which every $|k_2|+|k_3|<\delta$ admits a four-dimensional regular
non-polynomial family. Choosing the two signs reaches
\begin{equation}
 (+,+,+),\quad(+,+,0),\quad(+,+,-),\quad(+,0,0),\quad
 (+,-,0),\quad(+,-,-),
 \label{np-analytic-all-signs}
\end{equation}
up to permutation. Compact surfaces of the appropriate genus are chosen
separately in each sector. Appendix~\ref{curvature-normalization} puts
their curvatures in the unit-sign convention without changing the physical
background. This proves sufficiency and completes the sign criterion.

For comparison, the compact regular classes explicitly reported by GMMW
\cite[sections 4, 5 and 8]{Gauntlett:2006ux} occupy only three of these
six sectors, as mentioned previously. 
In particular, the exact $(+,0,0)$ seed \eqref{np-reference-solution-functions} has unequal
functions $Q_2$ and $Q_3$ on its two flat factors, so it is not in the
specialised $KE_4\times KE_2$ subclass considered in section 4 of GMMW.

\subsection{Local flux coordinates and their physical range}
\label{np-analytic-flux}\label{np-analytic-sign-wedge}

The analytic family above is naturally parametrized by the four coefficients
$q=(A_2,B_2,A_3,B_3)$ after the endpoint equations have eliminated
$A_1,B_1$.  For holography, however, the more useful coordinates are the
magnetic periods.  Because radial translation and homogeneous scaling act
trivially on the ratios \eqref{np-fluxmap-ratios}, four independent flux
ratios are exactly the number one expects on the quotient radial family.
The remaining point is to check that the flux map really has rank four.

Differentiating \eqref{np-analytic-endpoint-F} gives
\begin{equation}
 D_q(A_1,B_1)\big|_0=
 \begin{pmatrix}
 -\tfrac18&\tfrac14&-\tfrac1{16}&-\tfrac14\\
 \tfrac1{16}&-\tfrac1{16}&-\tfrac1{16}&-\tfrac18
 \end{pmatrix}.
 \label{np-flux-J-dependent}
\end{equation}
Substituting these derivatives into the endpoint-period formulae gives, for
$\mathcal F=(x,y,u,v)$,
\begin{equation}
 D_q\mathcal F\big|_0=
 \begin{pmatrix}
 -\tfrac18&\tfrac18&-\tfrac1{18}&-\tfrac7{36}\\
 -\tfrac1{18}&\tfrac5{36}&-\tfrac18&-\tfrac14\\
 \tfrac1{144}&-\tfrac1{144}&-\tfrac1{144}&-\tfrac1{72}\\
 -\tfrac1{18}&\tfrac5{36}&-\tfrac1{24}&-\tfrac1{12}
 \end{pmatrix},
 \qquad
 \det D_q\mathcal F=-\frac1{248832}.
 \label{np-analytic-flux-J}
\end{equation}
Exact differentiation can be found in Appendix~\ref{np-exact-flux-jacobian}.
The nonzero determinant has a simple geometric meaning, i.e., after quotienting
the two radial redundancies, the flux ratios separate the nearby regular
solutions.  Equivalently, at fixed curvature parameters and fixed base
volumes, the map from quotient radial data to the four ratios is a local
analytic diffeomorphism.

The distinction between this local statement and a global description of
the flux image is important.  Equation \eqref{np-analytic-flux-J} implies
that the image contains an open neighbourhood of the reference flux point,
but it neither gives the size of that neighbourhood, nor says that every
algebraically admissible flux vector is realized.  It also does not identify
different compactification data. Changing a genus or a torus volume changes
the relation between the local periods $I_i,G_i^-$ and the physical
integers $N_i,M_i$.  Throughout the local-coordinate statement, the base
geometry and the volume factors $\nu_i$ are held fixed.

The exact reference periods are
\begin{equation}
 (I_1,I_2,I_3;G_1^-,G_2^-,G_3^-)
 =
 \left(
 -\frac94,-\frac3{16},-\frac3{16};
 0,-\frac3{16},\frac38
 \right),
 \qquad x=y=\frac1{12}.
 \label{np-reference-periods}
\end{equation}
Now turn on small nonzero curvatures $k_a$, $a=2,3$, while keeping
$k_1=\nu_1=1$.  At fixed genus, Gauss--Bonnet gives
$n_a=k_a\nu_a$, so that
\begin{equation}
 \frac{N_a}{N_1}
 =
 \frac{n_a}{k_a}
 \left(\frac1{12}+o(1)\right),
 \qquad a=2,3.
 \label{np-analytic-aspect-ratios}
\end{equation}
Thus, the perturbative neighbourhood has a characteristic physical
interpretation, namely, when the new curvatures are taken small at fixed topology,
the corresponding reference areas grow and the physical fibre-flux aspect
ratios become large.  Nearby non-polynomial seeds keep the local ratios
bounded away from zero, so this large-aspect-ratio behaviour is not a
peculiarity of the exact reference solution.

This is why the implicit-function neighbourhood says little about
arbitrary finite flux ratios. Curvature normalization does not change the
physical ratios, so normalizing a small curvature to unit magnitude does
not move the solution from the asymptotic region of flux space to a generic
finite-ratio point.  The analytic
construction proves open non-polynomial families in every allowed sign
sector and gives local flux coordinates there, but does not characterize the
full image of the flux map.  Section~\ref{flux-space} addresses that
separate problem directly by first reconstructing candidate endpoint data
from fluxes and then certifying the finite-ratio neighbourhood
$\mathcal B_+$ defined in \eqref{np-positive-flux-box} with an independent
interior argument.

\section{Flux reconstruction and a certified image patch}\label{flux-space}

\subsection{Rational reconstruction of candidate endpoints}
\label{np-flux-endpoint-inverse}

The local inverse-function result of section~\ref{np-analytic-flux}
can be complemented by an explicit rational inverse at the level of
endpoint data. The construction below recovers candidate endpoints from
the fluxes on a nondegenerate chart, although their regular filling remains a
separate boundary-value problem.

We fix $(k_i,\nu_i)$, an unshifted vector $(N_i;M_i)$ and a positive overall
scale $\fluxscale$.  We then define the local periods
\begin{equation}
 p_i=\frac{N_i}{\fluxscale\nu_i},\qquad
 g_i=\frac{M_i}{2\fluxscale\nu_j\nu_\ell},
 \label{np-inverse-local-periods}
\end{equation}
and the north-section combination
\begin{equation}
 g_i^+=g_i+k_jp_\ell+k_\ell p_j,\qquad
 S_k=\sum_{\rm cyc}k_i p_jp_\ell.
 \label{np-inverse-flux-input}
\end{equation}
The endpoint data of any regular solution obey the compatibility relation
\begin{equation}
 p\cdot g+S_k=0.
 \label{np-inverse-compatibility}
\end{equation}
We introduce
\begin{equation}
 \mathsf L_p=
 \begin{pmatrix}
 0&p_3&p_2\\
 p_3&0&p_1\\
 p_2&p_1&0
 \end{pmatrix},
 \qquad
 \eta=\mathsf L_p^{-1}g^+,\qquad
 \Delta_g=\eta\cdot g.
 \label{np-inverse-matrix}
\end{equation}
On the non-degenerate branch
\begin{equation}
 p_1p_2p_3\,\Delta_g\,S_k\neq0,
 \label{np-inverse-nondegenerate}
\end{equation}
the candidate interval length and the six pole ratios are
\begin{align}
 \delta&=\frac{2S_k}{\Delta_g},
 \qquad r_-=0,\quad r_+=\delta,
 \label{np-inverse-length}\\
 z_i^-&=-p_i-\frac{\delta}{2}\eta_i,\qquad
 z_i^+=p_i+\frac{\delta}{2}(k_i-\eta_i),
 \label{np-inverse-poles}
\end{align}
where $z_i^\pm$ is the candidate value of $H'_\pm/Q_i^\pm$.
The quadratic coefficients then follow from
\begin{equation}
 d_i=-z_j^-z_\ell^-,
 \qquad
 c_i=\frac14K_i\delta
 -\frac{z_j^+z_\ell^+-z_j^-z_\ell^-}{\delta}.
 \label{np-inverse-parameters}
\end{equation}
Every quantity in \eqref{np-inverse-length}--\eqref{np-inverse-parameters}
is rational in the prescribed fluxes once $(k_i,\nu_i,\fluxscale)$ are fixed.

This derivation is short enough to make the meaning of the formulae
transparent. I.e., let
\begin{equation}
 z_i^\pm=\frac{H'_\pm}{Q_i^\pm},
 \label{np-inverse-zdef}
\end{equation}
The fibre-period formula \eqref{np-general-fibre-periods} then gives
\begin{equation}
 z_i^+-z_i^-=2p_i+\frac12k_i\delta.
 \label{np-inverse-zdifference}
\end{equation}
The south-section periods, together with the quadratic interpolation of
$Q_i$, give
\begin{equation}
 g^+=-\frac2\delta\,\mathsf L_p(z^-+p).
 \label{np-inverse-linear-system}
\end{equation}
Thus, $z^-+p=-(\delta/2)\eta$, which is the first equation in
\eqref{np-inverse-poles}.  Supersymmetry at the south pole gives
$z^-\cdot g=0$, and hence
\begin{equation}
 p\cdot g+\frac{\delta}{2}\Delta_g=0.
 \label{np-inverse-delta-step}
\end{equation}
Using \eqref{np-inverse-compatibility} yields
\eqref{np-inverse-length}.  Equation \eqref{np-inverse-zdifference} then
gives $z^+$.  Finally, the endpoint equations imply
\begin{equation}
 Q_i^\pm=4z_j^\pm z_\ell^\pm,\qquad
 H'_\pm=4z_1^\pm z_2^\pm z_3^\pm,
 \label{np-inverse-Q-slopes}
\end{equation}
and interpolation of the quadratics gives
\eqref{np-inverse-parameters}.  Substitution recovers $I_i=p_i$ and
$G_i^-=g_i$ identically, using only endpoint data.

The arbitrary positive scale $\fluxscale$ simply rescales all $p_i$ and
$g_i$ together, giving the homogeneous radial scaling
\eqref{np-finite-scale}, leaving the quotient endpoint shape unchanged.  Conversely, a vanishing denominator
in \eqref{np-inverse-nondegenerate} does not, by itself, signal a singular
geometry. It means only that this particular eliminated form of the
endpoint equations is not a good chart, and the uneliminated equations must
be used.

The reconstruction also gives simple algebraic regularity tests.  In the
chosen orientation,
\begin{equation}
 \delta>0,\qquad z_i^->0,\qquad z_i^+<0
 \label{np-inverse-sign-tests}
\end{equation}
are required.  Writing $Q_i^\pm=4z_j^\pm z_\ell^\pm$, the quadratic
identity \eqref{np-boundary-Q-square}, after rescaling the interval to
$[0,\delta]$, gives
\begin{equation}
 Q_i>0\text{ on }[0,\delta]
 \quad\Longleftrightarrow\quad
 \sqrt{Q_i^-}+\sqrt{Q_i^+}>
 \delta\sqrt{K_i}\qquad(K_i>0).
 \label{np-inverse-Q-test}
\end{equation}
For $K_i\leq0$, positive endpoint values already imply positivity on the
whole interval.  A further necessary condition is the positivity of the
endpoint volume,
\begin{equation}
 \mathcal I_{\rm end}
 =-16p_1p_2p_3-4\delta S_k>0,
 \label{np-inverse-volume-test}
\end{equation}
as follows from the endpoint identity \eqref{np-endpoint-volume-flux}.
These conditions define an algebraically admissible set.  They do not
replace the interior requirements $H>0$ and $D<0$.

For an exact example of the inverse, consider the
fully positive representative at the centre of the certified box
$\mathcal B_+$ defined in \eqref{np-positive-flux-box}.
Table~\ref{examples-all-chambers} in section~\ref{explicit-representatives} uses the unshifted convention. For the
fully positive representative the vector is
\begin{equation}
 \Gamma=(-1098,-3294,-3721;\,10248,10248,-1368),
 \qquad k_i=\nu_i=1.
 \label{np-inverse-positive-unshifted}
\end{equation}
Choosing $\fluxscale=1098$ gives
\begin{equation}
 p=\left(-1,-3,-\frac{61}{18}\right),\qquad
 g=\left(\frac{14}{3},\frac{14}{3},-\frac{38}{61}\right).
 \label{np-inverse-positive-pg}
\end{equation}
For these data,
\begin{align}
 S_k&=\frac{149}{9},&
 g^+&=\left(-\frac{31}{18},\frac5{18},-\frac{282}{61}\right),\notag\\
 \eta&=\left(\frac{118}{183},\frac{164}{61},-\frac{133}{54}\right),&
 \Delta_g&=\frac{28147}{1647},
 \label{np-inverse-positive-intermediate}
\end{align}
so that
\begin{equation}
 \delta=\frac{54534}{28147},
 \label{np-inverse-positive-delta}
\end{equation}
and
\begin{align}
 z^-&=\left(
 \frac{10565}{28147},
 \frac{11133}{28147},
 \frac{69662}{12063}\right),\notag\\
 z^+&=\left(
 -\frac{18462}{28147},
 -\frac{130482}{28147},
 -\frac{8662}{253323}\right).
 \label{np-inverse-positive-z}
\end{align}
The signs are exactly those required in
\eqref{np-inverse-sign-tests}.  After the homogeneous rescaling that sends
the interval to $[0,1]$, \eqref{np-inverse-parameters} gives
\begin{align}
 (c_1,c_2,c_3)
 &=\left(
 \frac{267083}{327204},
 \frac{268787}{327204},
 -\frac{577125}{1108858}\right),\notag\\
 (d_1,d_2,d_3)
 &=\left(
 -\frac{4944289}{8125566},
 -\frac{42228305}{73130094},
 -\frac{13068905}{330439684}\right).
 \label{np-inverse-positive-cd}
\end{align}
These are the exact centre data used in
appendix~\ref{np-positive-certificate}. More importantly, the inverse has
already fixed the candidate interval, pole slopes and all six quadratic
coefficients before the differential equation is solved. The remaining
question is interior existence, which is handled in the next subsection
uniformly in a neighbourhood of this example.

\subsection{A finite-ratio box in the fully positive sector}
\label{np-positive-flux-patch}
The rational inverse separates the algebraic endpoint problem from the
interior boundary-value problem. To show that the physical flux image also
contains a genuinely finite-ratio region, appendix~\ref{np-positive-certificate}
constructs explicit sub and supersolutions uniformly over the box below.
Their residuals have strict opposite signs, the associated semilinear
operator is monotone throughout the resulting tube, and a separate endpoint
identity gives $D<0$ in the fully positive sector. The argument establishes an explicit inner patch of the physical flux image.

For $k_i=\nu_i=1$, the four coordinates are physical ratios,
\begin{equation}
 f=(x,y,u,v)=\left(\frac{N_2}{N_1},\frac{N_3}{N_1},
                         \frac{M_1}{2N_1},\frac{M_2}{2N_1}\right).
 \label{np-positive-physical-ratios}
\end{equation}
The fully positive representative has
$f_*=(3,61/18,-14/3,-14/3)$ and $w_*=38/61$.
Every point in the closed box
\begin{equation}
 \mathcal B_+=\{f:\|f-f_*\|_\infty\leq10^{-3}\},\qquad
 w=-\frac{u+xv+xy+y+x}{y},
 \label{np-positive-flux-box}
\end{equation}
is realized by a smooth regular non-polynomial solution.
For a compatible unshifted vector with $N_1<0$ and ratios in
$\mathcal B_+$ of \eqref{np-positive-flux-box}, the scale $\fluxscale>0$ recovers all six fluxes. Because $p_1(TM_8)/4$ is integral in the present topology, these are
equivalently the compatible shifted-integral M-theory flux vectors of the
same orientation.

Appendix~\ref{np-positive-certificate} proves this statement with
polynomial barriers after normalizing the radial interval to $[0,1]$. The
uniform bounds include
\begin{gather}
 Q_1>0.15,\qquad Q_2>0.015,\qquad Q_3>0.156,
 \label{np-positive-Q-bounds}\\
 0.050r(1-r)<H(r)<0.477r(1-r),\qquad
 D(0)<-7.77,\quad D(1)<-1.82.
 \label{np-positive-profile-bounds}
\end{gather}
The strict bounds on $H$ apply for $0<r<1$.
The endpoint-sheet identity then gives $D<0$ in the interior, and the
barriers give uniqueness inside the tube. Polynomial solutions are excluded
separately by the quartic compatibility obstruction in
appendix~\ref{np-positive-certificate}. Thus, although this inner estimate does not locate the
boundary of the flux image, $\mathcal B_+$ is an explicit patch of the physical flux image, obtained independently of the
small-curvature construction.

\subsection{The integral lattice and quantized families}
\label{np-shifted-lattice}
The compact space is $M_8=\mathbb P(\mathcal O\oplus\mathcal L)$ over
$B_6$, with the anti-canonical twisting convention
\begin{equation}
 \ell=c_1(\mathcal L)=2\sum_i n_i\omega_i,\qquad
 \omega_i=\frac{J_i}{4\pi\nu_i},\qquad \int_{C_i}\omega_i=1.
 \label{np-lattice-twist}
\end{equation}
For the projective generator $u_B$, choose
$u_B(u_B-\pi^*\ell)=0$ and $u_B|_-=0$. The projective-bundle formula
gives
\begin{equation}
 H^4(M_8,\mathbb Z)=\pi^*H^4(B_6,\mathbb Z)
                \oplus u_B\pi^*H^2(B_6,\mathbb Z),\qquad
 \Gamma=\sum_{\rm cyc}M_i\pi^*(\omega_j\omega_\ell)
           +\sum_iN_i u_B\pi^*\omega_i.
 \label{np-lattice-Gamma}
\end{equation}
These groups are torsion-free. Mixed classes built from one-forms on
different positive-genus factors have zero periods for this flux ansatz
and for the characteristic-class shift. The six displayed generators
therefore suffice for its quantization. Note that, at the north section,
$u_B|_+=\pi^*\ell$.  Restricting the cohomology class $\Gamma$ therefore
gives the purely topological relation
\begin{equation}
 M_i^+=M_i+2(n_jN_\ell+n_\ell N_j),
 \label{np-lattice-pole-relation}
\end{equation}
which is fixed by the projective-bundle topology, i.e., it does not use
the supersymmetry equations. In particular, the factor of two is forced by
$\ell=2\sum_i n_i\omega_i$ and is not an independent normalization
choice.

Fluxes that arise from the GMMW endpoint system obey one further condition,
\begin{equation}
 \sum_iN_iM_i+2\sum_{\rm cyc}n_iN_jN_\ell=0.
 \label{np-lattice-compatibility}
\end{equation}
This is the physical version of
\eqref{np-general-flux-compatibility}, and should not be confused with
\eqref{np-lattice-pole-relation}. The former is a quadratic compatibility
constraint imposed by the endpoint/supersymmetry equations, whereas the
latter simply relates the restrictions of one and the same cohomology
class to the two sections.  In particular, an arbitrary integral class on
the projective bundle satisfies \eqref{np-lattice-pole-relation}, but need
not satisfy \eqref{np-lattice-compatibility}.
On a nondegenerate chart with the base data fixed, the single
constraint leaves five continuous flux parameters, corresponding to four
quotient radial shapes and the overall scale $\fluxscale$. After
quantization, however, the allowed vectors are integral points on this
quadratic constraint, not five freely chosen integers.

We orient the projective fibre so that $\pi_*u_B=1$. With the relation
$u_B(u_B-\pi^*\ell)=0$, the complex vertical tangent line then has
$c_1(T_\pi)=2u_B-\pi^*\ell$, whose integral over the fibre is $2$.
Its restrictions to the south and north sections are $-\ell$ and $+\ell$,
respectively. These are the normal Chern classes used later in
section~\ref{np-localization-reproducible}.
Squaring and using the projective relation yields
$p_1(T_\pi)=\pi^*\ell^2$. The tangent sequence and $p_1(TB_6)=0$ give
\begin{equation}
 p_1(TM_8)=\pi^*[p_1(TB_6)+\ell^2],\qquad
 \frac14p_1(TM_8)=2\sum_{i<j}n_in_j\pi^*(\omega_i\omega_j).
 \label{np-lattice-shift}
\end{equation}
The total space is spin. Every
orientable Riemann surface is spin, hence $w_2(TB_6)=0$.  The vertical
complex line has $c_1(T_\pi)=2u_B-\pi^*\ell$, so
$w_2(T_\pi)=\pi^*\ell\bmod2$.  Since
$\ell=2\sum_i n_i\omega_i$, one obtains
$w_2(TM_8)=0$, and, thus, $M_8$ itself is spin (although the spin structure need not be
unique when $H^1(M_8,\mathbb Z_2)\neq0$), and the usual M-theory shifted
quantization law is well defined.

Moreover, in the present topology, the shift itself is integral.
\eqref{np-lattice-shift} shows that $p_1(TM_8)/4$ is an integral class, and, consequently, the Witten condition \cite{Witten:1996}
$\Gamma_{\rm sh}=\Gamma-p_1(TM_8)/4\in H^4(M_8,\mathbb Z)$ is equivalent
here to ordinary integrality of the six unshifted periods.  There are no
half-integral magnetic periods in these examples, so the shift simply
translates the integer labels by
$(0,0,0;2n_2n_3,2n_3n_1,2n_1n_2)$.  We retain both shifted and unshifted
vectors because the distinction is useful when comparing M-theory
conventions and anomaly formulae.

The quantization statements in this paper concern the magnetic class
$\Gamma$ and its characteristic-class shift. The electric coefficient
$g_7(r)$ in \eqref{np-gmmw-fourform} is fixed by the supersymmetry equations
once the background is specified, and a nonzero value of this coefficient does
not by itself identify a conserved M2 charge. At two-derivative order the
local Page form is
\begin{equation*}
 \mathcal P_7=*_{11}G_4+\tfrac12 C_3\wedge G_4,
\end{equation*}
which is closed on shell where $C_3$ is defined. Its compact periods require
the patching of $C_3$ and a specification of any flat $C$-field data, and
can change under large gauge transformations \cite{Marolf:2000}.
The absence of an independent electric coefficient in the reduced equations
therefore does not by itself establish Page-charge quantization.

The projective-bundle formula gives
$H^7(M_8,\mathbb Z)=u_B\pi^*H^5(B_6,\mathbb Z)$. In the fully positive
sector, $B_6=(S^2)^3$ and this group vanishes, so there are no independent
compact seven-cycle Page periods. Note that bases of positive genus can have
nontrivial seven-cycles, but we do not evaluate their Page periods or classify
flat $C$-field data here. Consequently, the quantized families constructed
below are regular leading-order supergravity backgrounds satisfying the
shifted magnetic quantization law. 

Choose rational small nonzero curvatures and rational torus volumes in the
analytic construction. Then the volume factors are rational, and rational
points are dense in the open physical flux image supplied by
\eqref{np-analytic-flux-J} and the non-polynomial subset is open. Clearing
the denominators with some $\fluxscale_0>0$ gives an integral unshifted class
$\Gamma_0$. Every $\fluxscale=m\fluxscale_0$, $m\in\mathbb Z_{>0}$, then gives
$m\Gamma_0-p_1(TM_8)/4\in H^4(M_8,\mathbb Z)$.
Curvature normalization preserves this tower, establishing infinitely many
quantized non-polynomial backgrounds in every allowed sector, and the
certified box $\mathcal B_+$ of \eqref{np-positive-flux-box} adds
rational shapes at finite ratios.

Within the analytic coordinate neighbourhood, fixed flux ratios determine
the quotient radial shape locally, with the base geometry and its volumes
held fixed. Along an integral flux ray, the leading central charge scales
as $m^3$ and the gravitational anomaly as $m$.

\section{Representatives and profile deviations}\label{explicit-representatives}
\label{np-profile-deviation}

Table~\ref{examples-all-chambers} gives independently constructed targets
in the six allowed non-flat sectors, namely, the fully positive row lies at the
centre of the certified box $\mathcal B_+$ of
\eqref{np-positive-flux-box}, whilst the other finite-ratio targets are supported
by numerical boundary-value calculations, and the analytic sign-sector result
does not place them inside its perturbative neighbourhood.

To measure the non-polynomial correction, put
$\Delta_r=r_+-r_-$ and $t=(r-r_-)/\Delta_r$. The unique endpoint-matching cubic and its
relative uniform error are
\begin{equation}
 H_3=\Delta_r t(1-t)[(1-t)H'_- -tH'_+],\qquad
 \varepsilon_3=\frac{\|H-H_3\|_\infty}{\|H\|_\infty}.
 \label{np-profile-cubic-deviation}
\end{equation}
When all curvatures are nonzero, a polynomial solution must be quartic.
We also report
\begin{equation}
 \varepsilon_4=\min_{C\in\mathbb R}
   \frac{\|H-H_3-Ct^2(1-t)^2\|_\infty}{\|H\|_\infty}.
 \label{np-profile-quartic-deviation}
\end{equation}
If a curvature vanishes, polynomial degrees above three are excluded. Both deviations are invariant under the radial freedoms and curvature
normalization. They compare profiles with the same endpoint data. The
comparison polynomials need not solve the master equation, and since the
metric also involves derivatives of $H$, these are not metric-error bounds.

\begin{table}[htbp]
\centering\footnotesize
\setlength{\tabcolsep}{4pt}\renewcommand{\arraystretch}{1.25}
\begin{tabular}{@{}c c r c c@{}}
\toprule
Signs & Unshifted flux vector & $\cL-\cR$ & $\varepsilon_3$ & $\varepsilon_4$\\
\midrule
$(+,+,+)$ & $\begin{array}{c}(-1098,-3294,-3721;\\10248,10248,-1368)\end{array}$
 & $16226$ & $7.62\times10^{-2}$ & $1.21\times10^{-2}$\\
$(+,+,0)$ & $(-10,-11,-3;10,10,-28)$
 & $6$ & $4.85\times10^{-7}$ & n/a\\\
$(+,+,-)$ & $\begin{array}{c}(-361,-378,-228q_g;\\-6q_g,-855q_g,1708)\end{array}$
 & $-1022q_g$ & $3.31\times10^{-4}$ & $1.75\times10^{-8}$\\
$(+,0,0)$ & $\begin{array}{c}(-13104,-4536,-819;\\728,-1664,6640)\end{array}$
 & $0$ & $2.83\times10^{-4}$ & n/a\\\
$(+,-,0)$ & $\begin{array}{c}(-253,-244q_g,-291;\\-100q_g,409,-274q_g)\end{array}$
 & $-582q_g$ & $9.10\times10^{-4}$ & n/a\\\
$(+,-,-)$ & $\begin{array}{c}(-149,-65q_g,-3326q_h;\\-2999q_g q_h,3967q_h,-117q_g)\end{array}$
 & $-6484q_g q_h$ & $2.48\times10^{-2}$ & $5.39\times10^{-5}$\\
\bottomrule
\end{tabular}
\caption{Representatives in the allowed sectors. A positive, zero or
negative entry denotes $S^2$, $T^2$ or a higher-genus surface, respectively.
Here $q_g=g-1$ and $q_h=h-1$. Fluxes use the order
$(N_1,N_2,N_3;M_1,M_2,M_3)$, and the displayed vectors are unshifted,
matching the flux ratios and localization formulae used in the text.
Profile deviations are numerical estimates. The positive-sector box
$\mathcal B_+$ of \eqref{np-positive-flux-box} is certified independently
by appendix~\ref{np-positive-certificate}.}
\label{examples-all-chambers}
\end{table}

Table~\ref{examples-all-chambers} also shows why genericity should not be
read as a statement about the size of the deformation. Namely, the $(+,+,-)$
representative has $\varepsilon_4=1.75\times10^{-8}$ and is numerically
extremely close to an endpoint-matching quartic, whereas the fully positive
representative has $\varepsilon_4=1.21\times10^{-2}$. These quantities
measure profile distance only. A tiny nonzero value is fully compatible
with lying off the polynomial locus, and by itself is not a rigorous
polynomial-exclusion test for the individual numerical representative.

The deviations in Table~\ref{examples-all-chambers} can be reproduced from
its unshifted flux vectors and the volume factors in
appendix~\ref{np-representative-normalizations}, which gives the
endpoint reconstruction in a common unit-interval gauge. The profiles
were obtained by numerical collocation, with an independent two-sided
Frobenius shooting check of the $(+,+,0)$ benchmark. The endpoint
reconstruction and regularized boundary conditions are recorded in
appendix~\ref{np-representative-normalizations}.

The $(+,+,0)$ benchmark admits an exact polynomial-exclusion test,
independent of numerical profile comparisons or elimination of cubic
branches, as a polynomial solution in this sector has degree at most three.
Indeed, for a polynomial of degree $n\geq4$, the coefficient of
$r^{2n-2}$ in $2HH''-(H')^2$ is $n(n-2)h_n^2$, and neither $HR$ nor
$P/4$ can cancel it, since $R$ is linear and $P$ has degree at most four.
The two prescribed zeros and endpoint slopes therefore determine the
only possible polynomial filling, namely the cubic $H_3$ in
\eqref{np-profile-cubic-deviation}.

For the unshifted benchmark
$\Gamma=(-10,-11,-3;10,10,-28)$, take $(k_1,k_2,k_3)=(1,1,0)$,
$\nu_1=\nu_2=1$, and leave the torus volume $\nu_3>0$ arbitrary.
Set the temporary reconstruction scale $\fluxscale_0=10$ and use
\eqref{np-inverse-matrix}--\eqref{np-inverse-parameters} followed by
the unit-interval normalization in
appendix~\ref{np-representative-normalizations}. Exact rational
reconstruction gives
\begin{equation}
 \delta_0=\frac{8316}{17209},\qquad
 \beta=\frac{221}{594\nu_3},\qquad
 B_1=B_2=-\frac{662}{297\nu_3},
 \label{np-benchmark-exact-cubic-data}
\end{equation}
where $H_3=t(1-t)(\alpha+\beta t)$ and $B_1,B_2$ are the
coefficients of $t$ in the normalized linear functions $Q_1,Q_2$.
Substitution of the resulting cubic into the master equation, expressed
in the same unit-interval coordinate $t$, cancels the constant through
cubic coefficients by the endpoint equations. The $t^4$ coefficient
then gives the exact identity
\begin{align}
 \mathcal E_H[H_3]
 &=t^2(1-t)^2\left(3\beta^2+\beta(B_1+B_2)
                   +\frac14 B_1B_2\right)\notag\\*
 &=-\frac{49}{39204\nu_3^2}\,t^2(1-t)^2\neq0
 \qquad(0<t<1).
 \label{np-benchmark-exact-cubic-residual}
\end{align}
Thus, no polynomial solution of any degree has these endpoint data.
The coefficient is nonzero for every $\nu_3>0$, so the conclusion does
not depend on the rounded torus-volume entry in
appendix~\ref{np-representative-normalizations}. Together with the
regular numerical filling, this gives a direct obstruction for the
benchmark rather than merely a statement about generic cubic branches.

As a complementary flux-space diagnostic, on the regular generic cubic
branch of the $(+,+,0)$ sector, set
$\varrho=N_2/N_1$, $a=1+M_2/N_3$ and $b=1+M_1/N_3$.
Eliminating the cubic coefficients gives the necessary relation
\begin{equation}
 \widehat{\mathcal R}_{\rm poly}
 =b^3-b-\varrho^2(a^3-a)+2\varrho ab(b-a)=0,
 \label{np-polyflux-Rhat}
\end{equation}
and the denominator-cleared obstruction is homogeneous of degree five,
$\mathcal R_{\rm poly}=N_1^2N_3^3\widehat{\mathcal R}_{\rm poly}$.
This is the same quantized $(+,+,0)$ geometry used as the numerical
non-polynomial example in \cite{CompanionLocalisation}, up to the cyclic
ordering of the flux entries. For its unshifted fluxes,
\begin{equation}
 \varrho=\frac{11}{10},\qquad a=b=-\frac73,\qquad
 \mathcal R_{\rm poly}=-5880,\qquad
 \widehat{\mathcal R}_{\rm poly}=\frac{98}{45}.
 \label{np-polyflux-benchmark}
\end{equation}
The exceptional prefactor in that elimination,
$ab\varrho^6(a^2-1)(b^2-1)(ab-1)(3ab+1)(a\varrho+b)^4$,
is nonzero here. The flux obstruction provides an independent diagnostic
within its specified branch, while
\eqref{np-benchmark-exact-cubic-residual} is the complete polynomial
exclusion for the reconstructed benchmark endpoints. It is worth noticing that neither test here
interprets the small numerical profile deviation as a rigorous error
bound.

\subsection{A prescribed mixed-curvature target}\label{nonpoly-constructed}
The endpoint reconstruction can also be used by choosing flux ratios
before solving the interior problem. In the $(1,-1,1)$ chamber, take
\begin{equation}
 (x,y,u,v)=\left(\frac7{11},\frac{21}{20},\frac1{120},-\frac{71}{30}\right),
 \qquad w=\frac{547}{462}.
 \label{np-constructed-target}
\end{equation}
The fifth ratio follows from compatibility. With unit curvature magnitudes,
$\nu=(1,g-1,1)$, so for $g=2$ the unshifted integral vector is
\begin{equation}
 \Gamma=(-4620,-2940,-4851;-77,21868,-10940).
 \label{np-constructed-unshifted}
\end{equation}
The shifted pole periods are $(-75,21866,-10938)$. All fibre periods are
unchanged by the shift. Substitution in
\eqref{np-lattice-compatibility} verifies the integral constraint exactly.

The rational inverse fixes a candidate quotient endpoint shape for this
vector. The original endpoint solve, in a gauge fixing two of the reduced
constants, gives poles near $-0.583543797757156$ and
$-0.237198389970268$. Independent inward integration from the two poles
matches $(H,H')$ at approximately $10^{-12}$ after adjusting their second
derivatives. The computed profile has $H_{\max}\simeq0.6340636591$.
On the radial mesh the reconstructed coefficients obey
\begin{equation}
 3.9742\lesssim h_1\lesssim4.0785,\quad
 2.6330\lesssim h_2\lesssim3.3257,\quad
 3.3003\lesssim h_3\lesssim5.2317,
 \label{np-constructed-ranges}
\end{equation}
and $0.10551\lesssim\omega^6\lesssim0.11725$.
And, although they do not establish a validated surrounding neighbourhood, the matching and positivity checks support a regular filling of the chosen
flux target. The explicit neighbourhood proof
in this paper is the fully positive box $\mathcal B_+$ of
\eqref{np-positive-flux-box}.

For these data, the two possible leading quartic coefficients are
$-1/4$ and $-1/8$. Solving the higher polynomial coefficients leaves
numerically nonzero residuals $-13.6814495655356$ and
$5.07167170021773$, respectively. This example illustrates the two stages of the flux problem, i.e., exact candidate
endpoint reconstruction followed by a separate interior calculation.

\section{Endpoint observables and localization}\label{endpoint-observables}

\subsection{The warped volume}

The same cancellation that makes the magnetic periods endpoint quantities
also occurs in the warped volume. We can make that cancellation transparent by writing the factors separately as follows.

On the regular branch, $H>0$, $D<0$ and $Q_i>0$ and the unwarped
eight-dimensional volume form is
\begin{equation}
 \vol_8=
 h_1h_2h_3\sqrt{f_3f_4}\,
 J_1\wedge J_2\wedge J_3\wedge dr\wedge D\psi.
 \label{np-general-vol8}
\end{equation}
The reconstruction formulae give
\begin{equation}
 h_1h_2h_3=\frac{(-D)^3}{64P},\qquad
 \sqrt{f_3f_4}=\frac{-D}{4\sqrt P},\qquad
 \omega^9=\frac{8P^{3/2}}{(-D)^3}.
 \label{np-general-volume-factors}
\end{equation}
Multiplying these three expressions cancels every explicit power of $P$
and all but one power of $D$, i.e., 
\begin{equation}
 \omega^9\vol_8=-\frac D{32}
 J_1\wedge J_2\wedge J_3\wedge dr\wedge D\psi.
 \label{np-general-volume-density}
\end{equation}
Notice that no polynomial assumption enters this reduction. With
$e^{\lambda}=L_{\!A}\omega$ and $16\pi G_{11}=(2\pi)^8\ellp^9$,
the unit-radius Brown--Henneaux formula \cite{BrownHenneaux:1986}
(reproduced in equation~(4.21) of \cite{BenettiGenoliniEtAl:2024}
and equation~(2.25) of \cite{CompanionLocalisation}) gives
\begin{align}
 c_{\rm sol}^{(0)}
 &=\frac{3L_{\!A}^9}{2G_{11}}\int_{M_8}\omega^9\vol_8
 =\frac{3L_{\!A}^9}{2G_{11}}
   \frac{(2\pi)(4\pi)^3}{32}V\mathcal I\notag\\*
 &=\frac38\fluxscale^3 V\mathcal I,\qquad
 V=\nu_1\nu_2\nu_3,\qquad
 \mathcal I=-\int_{r_-}^{r_+}D\,dr.
 \label{np-general-central-endpoint}
\end{align}

The last integral is itself an endpoint observable.  Using
$D=2H''+R$ and
\[
 R=-\sum_i k_i(K_i r^2-4c_i r-4d_i)
\]
gives, in the translated gauge $r_-=0$, $r_+=\delta$,
\begin{equation}
 \mathcal I_{\rm end}
 =-2(H'_+-H'_-)
 -4\delta\sum_i k_i d_i
 -2\delta^2\sum_i k_i c_i
 +k_1k_2k_3\delta^3.
 \label{np-general-I-endpoint}
\end{equation}
For a regular solution, $\mathcal I_{\rm end}=\mathcal I$.  When
$R_\pm\neq0$, the differentiated pole condition may be used to eliminate
the slopes in favour of $P'_\pm/R_\pm$. When $R$ vanishes at a pole, one
simply keeps the slope form above, and, thus, no observable becomes singular
merely because the auxiliary ratio $P'/R$ is ill-defined at an endpoint.

The endpoint formula fixes the candidate leading central charge before
the interior profile is known. The comparison with localization below
uses these same endpoint data.

\subsection{The rational flux formula}\label{np-localization-reproducible}
We define the unshifted classes on the base by
\begin{equation}
 \mathfrak n=\sum_iN_i\omega_i,\qquad
 f_- =\sum_{\rm cyc}M_i\omega_j\omega_\ell,\qquad
 f_+=f_-+\ell\mathfrak n,
 \label{np-loc-n-fminus}
\end{equation}
where products denote wedge products. On the three-dimensional span of the
$\omega_i$, multiplication by $\mathfrak n$ is
\begin{equation}
 \mathsf L_{\mathfrak n}=\begin{pmatrix}0&N_3&N_2\\N_3&0&N_1\\N_2&N_1&0\end{pmatrix},
 \qquad a_+=\mathsf L_{\mathfrak n}^{-1}f_+.
 \label{np-loc-Lefschetz-matrix}
\end{equation}
We then set
\begin{equation}
 \mathsf I_3=\int_{B_6}\mathfrak n^3,\qquad
 \mathsf I_\pm=\int_{B_6}\mathfrak n f_\pm,\qquad
 J=\int_{B_6}a_+f_-,
 \label{np-loc-invariants}
\end{equation}
with the flux constraint $\mathsf I_++\mathsf I_-=0$.
On the branch $N_1N_2N_3J\neq0$, localization then gives
\begin{equation}
 c_{\rm loc}^{(0)}=-\mathsf I_3
                         +3\frac{\mathsf I_+\mathsf I_-}{J}.
 \label{np-loc-master-central}
\end{equation}
Notice that this equation uses the unshifted periods, whereas the characteristic-class shift enters flux quantization separately.

It is worth recording the specialized fixed-section data needed for the derivation.
The broader equivariant construction is developed in
\cite{BenettiGenoliniEtAl:2024,CompanionLocalisation}, with the central-charge
polyform and its normalization in equations~(4.20)--(4.21) of
\cite{BenettiGenoliniEtAl:2024}. We use the fixed-section formulas we derived in
section~4.2 of \cite{CompanionLocalisation}, particularly
its equations~(4.24)--(4.36) and (4.41)--(4.46), with $\chi$ replacing the class
called $x$ there and $f_\pm$ replacing $\mathfrak m_\pm$.
Here $S_{\rm loc}$ is the complex scalar Killing-spinor bilinear and
$P_{\rm loc}$ \footnote{We added the subscript to
distinguish $P_{\rm loc}$ from the radial polynomial
$P(r)=Q_1Q_2Q_3$.} is the complex one-form bilinear defined in equations~(2.4)
and~(2.5) of this same paper \cite{CompanionLocalisation}.  Notice that the GMMW sector belongs
to the subclass $S_{\rm loc}=P_{\rm loc}=0$.  In this subclass, the closed one-form
defining the equivariant scalar potential restricts to zero on a fixed
section, and since the pullback identifies $H^1(B_6)$ with $H^1(M_8)$, this form
is exact.  The same bilinear relation supplies the global electric
potential, including for bases with torus factors.

We write $\Lambda_p=(2\pi\ellp)^3$ and
$\beta_\pm=[e^{3\lambda}\omega_{\rm bil}|_\pm]$, with
$e^\lambda=L_{\!A}\omega$. Then choose the scalar-potential gauge $y_-=0$,
and put $\chi=-\pi\beta_-/\Lambda_p$ and
$\mathfrak s=\pi^2y_+/\Lambda_p$.
The normal weights are $\epsilon_\pm=\pm1$ and the normal Chern classes
are $\pm\ell$.  Specializing the equivariantly closed magnetic-flux and
supersymmetry polyforms to the two fixed sections gives the two relations
needed below,
\begin{equation}
 -\frac\pi{\Lambda_p}\beta_+=\chi+\mathfrak n-\mathfrak s\ell,
 \qquad \mathfrak n^2+2\chi\mathfrak n=2\mathfrak s f_+.
 \label{np-fixed-flux-constraints}
\end{equation}
The separate pole primitivity conditions are
$\chi f_-=0$ and $(\chi+\mathfrak n-\mathfrak s\ell)f_+=0$.
They follow directly here from $z_i^\pm=H'_\pm/Q_i^\pm$ and
\begin{equation}
 \sum_i z_i^\pm G_i^\pm
 =-\frac{P'_\pm}{16H'_\pm}-\frac{R_\pm}4=0,
 \label{np-fixed-primitivity-pole-check}
\end{equation}
as the fixed-section two-form is proportional to $\sum_i z_i^\pm J_i$.

The only further localization input is the fixed-section content of the
central-charge polyform.  In the present conventions, its components are
\begin{equation}
 (\Phi_6,\Phi_4,\Phi_2,\Phi_0)|_\pm
 =\left(\frac{\beta_\pm^3}{24},\,
 \frac{y_\pm\beta_\pm^2-\Lambda_p y_\pm^2f_\pm}{16},\,
 \frac{y_\pm^2\beta_\pm}{32},\,
 \frac{y_\pm^3}{192}\right).
 \label{np-fixed-volume-components}
\end{equation}
Thus, after expanding the inverse Euler class as
$(2\pi/\epsilon_\pm)\sum_{m=0}^3(-2\pi\ell)^m$, the fixed-point sum
gives
\begin{equation}
 c^{(0)}=-4\int_{B_6}
       [ (\chi+\mathfrak n)^3-\chi^3-3\mathfrak s^2\ell f_+ ].
 \label{np-fixed-central-before-elimination}
\end{equation}
The coefficient $-4$ uses the Brown--Henneaux calibration of the
central-charge polyform in the general localization formalism. The
algebra below tests the relative flux dependence at this normalization.

Equation~\eqref{np-fixed-flux-constraints} gives
$\chi=\mathfrak s a_+-\mathfrak n/2$. Primitivity then gives
$\mathfrak s J=\mathsf I_-/2=-\mathsf I_+/2$.
Using these relations in \eqref{np-fixed-central-before-elimination}
reduces it to $-\mathsf I_3+6\mathfrak s\mathsf I_+$ and proves
\eqref{np-loc-master-central}. At a zero denominator, we retain the
fixed-point constraints before elimination.

\subsection{Equality on the endpoint variety and the anomaly}

Here, by the endpoint variety, we mean the algebraic set of endpoint data satisfying
the equality conditions in \eqref{np-general-endpoint-two}, together with the
magnetic flux relations and the compatibility condition
\eqref{np-general-flux-compatibility}, without imposing the master equation
in the open interval.
The direct endpoint formula and the localized flux formula look rather
different, but on the endpoint variety they are the same algebraic
function.  This can be checked without using the interior master equation.
Set
\begin{equation}
 p_i=I_i,\qquad
 z_i^\pm=\frac{H'_\pm}{Q_i^\pm},\qquad
 E_\pm=\sum_i z_i^\pm G_i^\pm.
 \label{np-endpoint-Epm}
\end{equation}
The quantities $E_\pm$ are the pole primitivity combinations.
Expanding the quadratic interpolation of the $Q_i$ and using
$Q_i^\pm=4z_j^\pm z_\ell^\pm$ gives two useful identities,
\begin{equation}
 p\cdot G^-+S_k=\frac{E_+-E_-}{2},
 \qquad
 \mathcal I_{\rm end}
 =-16p_1p_2p_3-4\delta S_k+2\delta(E_++E_-).
 \label{np-endpoint-identity-certificates}
\end{equation}
For endpoint-compatible supersymmetric data, $E_+=E_-=0$.  The first
identity then reproduces the quadratic flux compatibility condition,
while the second gives
\begin{equation}
 \mathcal I_{\rm end}
 =-16p_1p_2p_3-4\delta S_k.
 \label{np-endpoint-volume-flux}
\end{equation}
This is the origin of the necessary volume test
\eqref{np-inverse-volume-test} in the inverse-flux construction.

The physical period dictionary gives
\begin{equation}
 \mathsf I_3=6\fluxscale^3Vp_1p_2p_3,\qquad
 \mathsf I_+=-\mathsf I_-=2\fluxscale^2VS_k,
 \label{np-endpoint-invariants-first}
\end{equation}
and the inverse endpoint identities give
\begin{equation}
 a_+=-\frac4\delta
 \sum_i\nu_i(z_i^-+p_i)\omega_i.
 \label{np-endpoint-a-plus}
\end{equation}
Using $E_-=0$ in $J=\int a_+f_-$ yields
\begin{equation}
 J=-\frac4{\fluxscale\delta}\mathsf I_-.
 \label{np-endpoint-J}
\end{equation}
Substitution into \eqref{np-loc-master-central} gives
\begin{align}
 c_{\rm loc}^{(0)}
 &=-6\fluxscale^3Vp_1p_2p_3
 +3\frac{(2\fluxscale^2VS_k)(-2\fluxscale^2VS_k)}
 {8\fluxscale VS_k/\delta}
 \notag\\
 &=-6\fluxscale^3Vp_1p_2p_3-\frac32\fluxscale^3V\delta S_k
 =\frac38\fluxscale^3V\mathcal I_{\rm end}.
 \label{np-endpoint-central-identity}
\end{align}
The last equality is \eqref{np-endpoint-volume-flux}. Thus, the
localization and endpoint expressions have identical flux dependence on
the endpoint variety, including the relative
$-\mathsf I_3+3\mathsf I_+\mathsf I_-/J$ structure. In a separate
analytic-background comparison in section~5.2 of
\cite{CompanionLocalisation}, equation~(5.32), we recover the known $T^2\times KE_4$
central charge from its Type IIB description.

These identities use the pole constraints, topology and magnetic periods,
not an interior solution. Endpoint observables can remain finite even as
the geometry becomes singular, as illustrated by the interval-collapse
limits discussed briefly in section~\ref{np-finite-boundary-classification}.

The $C_3\wedge X_8$ coupling fixes the gravitational anomaly
\cite{FreedHarveyMinasianMoore:1998,KrausLarsen:2005},
\begin{equation}
 \cL-\cR
 =-\frac14\int_{B_6}
 \mathfrak n\,[p_1(TB_6)+\ell^2]
 =-2\sum_{\rm cyc}N_i n_jn_\ell.
 \label{np-loc-product-anomaly}
\end{equation}
With the fibre orientation $\pi_*u_B=1$ used in
\eqref{np-lattice-Gamma}, one has $\pi_*\Gamma=\mathfrak n$, and hence
\begin{equation}
 \int_{M_8}\Gamma\wedge p_1(TM_8)
 =\int_{B_6}\mathfrak n\,[p_1(TB_6)+\ell^2]
 =8\sum_{\rm cyc}N_i n_jn_\ell.
 \label{np-anomaly-pushforward}
\end{equation}
The sign in \eqref{np-loc-product-anomaly} uses our
$\mathcal N=(2,0)$ boundary orientation. Reversing it exchanges
$c_{\mathrm L}$ and $c_{\mathrm R}$. Since $p_1(TM_8)/4$ is pulled back
from the base, $\pi_*\Gamma_{\rm sh}=\pi_*\Gamma$. The anomaly is
therefore unchanged when expressed using shifted, rather than unshifted
fluxes. It is linear in the fibre fluxes, as the characteristic class
is fixed by the compactification data, unlike the leading parity-even
central charge, which scales cubically along an integral flux ray.

For the unshifted benchmark
\begin{equation}
 (N_i;M_i)=(-10,-11,-3;\,10,10,-28),
 \qquad (n_1,n_2,n_3)=(1,1,0),
 \label{np-observable-benchmark}
\end{equation}
the invariants are
\begin{equation}
 \mathsf I_3=-1980,\qquad
 \mathsf I_+=126=-\mathsf I_-,
 \qquad
 J=\frac{17209}{165}.
 \label{np-observable-benchmark-invariants}
\end{equation}
Hence,
\begin{equation}
 c^{(0)}
 =1980
 +3\,\frac{126(-126)}{17209/165}
 =\frac{26215200}{17209},
 \qquad
 \cL-\cR=6.
 \label{np-observable-benchmark-values}
\end{equation}
The anomaly determines the difference of the two central charges, not the
full parity-even higher-derivative correction to their sum.

\section{Continuation and finite limits}\label{global-structure}

The open sets constructed above do not determine how the regular components
fit together globally. We can, nevertheless, say two things. A numerical
continuation connects several mixed-curvature representatives in the
coefficient space of the reduced ODE, and under explicit compactness
hypotheses a finite first loss of regularity is restricted to two boundary
mechanisms. The continuation is not a path at fixed compact topology or
through the integral flux lattice. Notice that neither of these statements is used in the
existence proof or in the certified positive-flux box $\mathcal B_+$ of
\eqref{np-positive-flux-box}.

\subsection{Numerical paths in the reduced parameters}
\label{nonpoly-continuation}

The reduced equations admit a numerical path joining the mixed-curvature
representatives used above.  It passes from the $(1,-1,-1)$ solution to the
zero-curvature sector and then, after varying the remaining reduced
parameters and permuting the last two factors, reaches the $(1,1,-1)$
representative.  Appendix~\ref{continuation-data} records the actual
continuation prescription and sampled nodes.  Here we use the path only as
evidence about connectedness in coefficient space.

This is a statement about the reduced equations, not a continuous family of
compact backgrounds.  Gauss--Bonnet gives
\begin{equation}
 \kappa\nu_3=1-g_3,
 \label{np-cont-topology}
\end{equation}
so crossing $\kappa=0$ changes the compact topology of $C_3$, while, at
fixed $g_3\neq1$, its area diverges as $\kappa\to0$.  Moreover, intermediate
continuation nodes are not required to obey integral flux quantization.
Thus, the calculation probes the local ODE solution space only.

Path dependence is visible already in a one-parameter slice.  Holding
the six reduced constants fixed eventually drives the two physical
endpoints together, whereas allowing the remaining parameters to vary
avoids that collision along the sampled path.  The numerical location of
the fixed-constant collision is given in appendix~\ref{continuation-data}.
Thus, a degeneration on a particular slice need not be a boundary of the
full regular component.

\subsection{Finite degeneration mechanisms}\label{phase-boundaries}
\label{np-finite-boundary-classification}

Consider a one-parameter family of regular solutions in a fixed radial
gauge.  Assume that the curvatures, the six reduced constants and the two
endpoints have finite limits.  If the limiting interval has positive length,
assume $C^3$ convergence on the closed interval and smooth convergence on
compact interior subintervals.  Under these hypotheses, a first loss of
regularity can occur only if the interval collapses, or if at least one $Q_i$
vanishes somewhere on the limiting closed interval.  The proof is given in appendix~\ref{finite-degenerations}, including the
exclusion of an independent first loss of $D<0$ while all $Q_i$ remain
positive.

For a collapse with $r_\pm\to r_{\rm c}$ and
$P_{\rm c}=P(r_{\rm c})>0$, the two physical endpoint sheets meet and hence
\begin{equation}
 P'_{\rm c}=R_{\rm c}=0.
 \label{np-boundary-collapse-condition}
\end{equation}
The endpoint observables nevertheless remain finite,
\begin{equation}
 I_i\longrightarrow-\frac{\sqrt{P_{\rm c}}}{2Q_i^{\rm c}},\qquad
 \mathcal I\longrightarrow2\sqrt{P_{\rm c}},\qquad
 Q_i^{\rm c}=Q_i(r_{\rm c})>0.
 \label{np-boundary-collapse-observables}
\end{equation}
Consequently, the limiting local flux ratios obey
\begin{equation}
 k_1+\frac{k_2}{x}+\frac{k_3}{y}=0,
 \label{np-collapse-flux-wall}
\end{equation}
and the physical ratios $X=N_2/N_1$, $Y=N_3/N_1$ satisfy
\begin{equation}
 n_1+\frac{n_2}{X}+\frac{n_3}{Y}=0.
 \label{np-collapse-physical-wall}
\end{equation}
Notice that this is a necessary boundary condition on the flux image, but not a statement
that every point on the wall is attained.  The two numerical collapses in
appendix~\ref{continuation-data} satisfy it at the precision shown there.
If all $k_i\geq0$ and at least one is positive, then
$R=-\sum_i k_iQ_i<0$, so such a pure collapse is impossible.

The second alternative is also restrictive.  Namely, if a single $Q_i$ develops
an isolated interior zero, while the other two remain positive, appendix~\ref{np-Q-pinch}
shows that the zero is necessarily double, that $H$ vanishes there with
$H'=0$, and that the remaining two curvature coefficients are both
positive, making the resulting metric singular at finite proper distance.
One two-dimensional base factor expands while the complementary internal
directions collapse, and the curvature diverges with local powers
derived in the appendix.  We therefore interpret a $Q_i$--$H$ pinch as a
boundary at which the classical compactification breaks down. Within the
present ansatz, we find no indication that it is a removable bolt, an
orbifold limit, a regular brane throat, or a source carrying an additional
independent conserved charge. We find that these conditions
are necessary only. No such pinch is realized numerically here, and whether such a boundary admits an
intrinsically M-theoretic resolution is a separate question we do not touch.

The classification concerns finite limits satisfying the stated convergence
hypotheses.  It does not cover escape to infinity in reduced parameter
space, loss of compactness of the profile, or noncompact ends of the regular
moduli space.  The completely flat family is instead a smooth exceptional
stratum, described in appendix~\ref{np-flat-corner}.

\section{Discussion}\label{discussion}

Apart from the completely flat exception, we find that a
positive-curvature factor is necessary and sufficient for regular compact
solutions in the present ansatz. Necessity follows from the endpoint no-go
argument of section~\ref{nonpoly-nogo}, while sufficiency comes from the
analytic families obtained by deforming the
$S^2\times T^2\times T^2$ reference solution into every allowed sign
sector. With the base geometry fixed, these families have four radial
shape parameters, and the polynomial condition cuts out a proper
analytic locus within them.

The flux-space picture is less complete. Near the analytic seed, four
magnetic flux ratios are genuine local coordinates and rational points give
infinitely many shifted-flux-quantized non-polynomial backgrounds after the
overall scale is cleared. At fixed genus, these neighbourhoods run to large
physical flux aspect ratios as the new curvatures become small, so they do
not describe generic finite-ratio fluxes.

At finite ratios, the rational endpoint inverse is useful because it settles
all of the algebra before one solves the differential equation. For  instance, for fixed
curvatures, topology and base volumes, a compatible flux vector determines
the candidate interval, pole ratios and six quadratic coefficients.
Orientation and the endpoint positivity tests can then be checked
immediately, and the remaining issue is genuinely an interior one, namely,  whether
there is a smooth profile with $H>0$ and $D<0$ between the poles.

The fully positive box $\mathcal B_+$ of \eqref{np-positive-flux-box}
gives one explicit finite-ratio answer. It is an inner patch of the
physical flux image, proved independently of the small-curvature
construction. The representatives in the other chambers remain
numerical examples, and we neither claim that they lie in the analytic
neighbourhood, nor that a validated continuation joins them to it.

The endpoint formulae simplify the observables as well. Both the magnetic
periods and the leading central charge are determined by pole data, and,
after the common Brown--Henneaux normalization is fixed, the localized and
direct expressions have the same flux dependence algebraically on the
endpoint variety. The gravitational anomaly is fixed separately by
the characteristic class and the magnetic flux.

These backgrounds retain the wrapped-M5 G-structure origin of the GMMW
sector and generally have a nonzero electric component of $G_4$. As
explained in section~\ref{np-shifted-lattice}, this field component alone
does not determine a conserved M2 Page charge. Along an
integral flux ray, the leading central charge grows cubically and the
gravitational anomaly linearly, as expected from the hierarchy of wrapped-M5
anomaly data
\cite{FreedHarveyMinasianMoore:1998,BenettiGenoliniEtAl:2023M5}. We do not
take this scaling to identify the microscopic SCFTs or to put all of the
backgrounds in one universality class, and doing a direct field-theory derivation of
the full flux-dependent anomaly polynomial would be very interesting.

The global geometry of the regular solution space also remains open. Under the
stated convergence assumptions, a finite first loss of regularity can only
occur through interval collapse or a $Q$ degeneration, and an independent
$D$ boundary and a generic same-sheet fold are excluded. Cross-sheet
collapse maps to the flux wall \eqref{np-collapse-physical-wall}, although
the limiting metric is singular.

A natural next problem would be continuation in physical flux coordinates at
fixed compactification data. Starting from the certified positive box
$\mathcal B_+$ would turn the sampled continuation problem into a
validated one. A path from that finite-ratio region to an analytic seed
would identify one connected piece of the physical flux image. Its maximal
extent, possible noncompact ends and the distribution of primitive
quantized rays remain unknown. Toric or non-K\"ahler bases would require a
different interior problem, rather than a straightforward enlargement of
the present endpoint analysis, but would also be most interesting.

\appendix
\section{Curvature normalization}\label{curvature-normalization}

Let $\lambda_i>0$, $\Lambda=\lambda_1\lambda_2\lambda_3$ and
$\mu_i=\Lambda/\lambda_i$. The simultaneous transformations
\begin{equation}
 \widehat k_i=\lambda_i k_i,\quad
 (\widehat c_i,\widehat d_i)=\mu_i(c_i,d_i),\quad
 \widehat H=\Lambda H,\quad
 (\widehat g_{C_i},\widehat J_i)=\lambda_i^{-1}(g_{C_i},J_i)
 \label{np-normalization-map}
\end{equation}
leave $r,\psi$ and the physical scale $\fluxscale$ fixed. The reconstruction
formulae give the following multiplicative weights.
\begin{center}
\begin{tabular}{@{}ll@{}}
\toprule
Quantities & Weight\\
\midrule
$k_i,h_i,g_{i+3},I_i$ & $\lambda_i$\\
$Q_i,c_i,d_i,g_i,G_i^\pm$ & $\mu_i$\\
$H,R,D,\mathcal I$ & $\Lambda$\\
$P$ & $\Lambda^2$\\
$g_{C_i},J_i,\nu_i$ & $\lambda_i^{-1}$\\
$f_3,f_4,\omega,\varphi,g_7$ & $1$\\
\bottomrule
\end{tabular}
\end{center}
For the electric coefficient, each term in the brackets of
\eqref{np-gmmw-electric-function} has weight $\Lambda$, cancelled by
$h_1h_2h_3$. Since $\widehat k_i\widehat J_i=k_iJ_i$, one can choose
the same circle-bundle connection, including its flat part and $2\pi$
fibre period. Thus, every term of the metric and four-form is unchanged,
\begin{equation}
 \widehat{ds_{11}^2}=ds_{11}^2,\qquad \widehat G_4=G_4.
 \label{np-normalization-fields}
\end{equation}
The integral forms $\omega_i=J_i/(4\pi\nu_i)$, twisting class and
projective bundle are the same. All physical periods, their
characteristic-class shifts and the holographic observables are therefore
invariant. Explicitly, $\nu_iI_i$ and $\nu_j\nu_\ell G_i$ have weight
one. Taking $\lambda_i=1/|k_i|$ for $k_i\neq0$, and $\lambda_i=1$
otherwise, normalizes the curvature signs. This equivalence preserves the
physical flux image of any existence construction.

\subsection{Normalization data for the numerical representatives}
\label{np-representative-normalizations}
The numerical profiles in Table~\ref{examples-all-chambers} can be
reconstructed using only the unshifted flux vectors in that table and the
curvatures and volume factors below, where rows are identified by curvature signs.
For negative curvature, $q_g=g-1$ and $q_h=h-1$. The displayed torus volumes
are the rounded numerical inputs, while all other entries are exact.

\begin{table}[htbp]
\centering\small
\begin{tabular}{@{}c c c@{}}
\toprule
Signs & $(k_1,k_2,k_3)$ & $(\nu_1,\nu_2,\nu_3)$\\
\midrule
$(+,+,+)$ & $(1,1,1)$ & $(1,1,1)$\\
$(+,+,0)$ & $(1,1,0)$ & $(1,1,0.04940596019)$\\
$(+,+,-)$ & $(1,1,-1)$ & $(1,1,q_g)$\\
$(+,0,0)$ & $(1,0,0)$ & $(1,1,1)$\\
$(+,-,0)$ & $(1,-1,0)$ & $(1,q_g,0.53388129901)$\\
$(+,-,-)$ & $(1,-1,-1)$ & $(1,q_g,q_h)$\\
\bottomrule
\end{tabular}
\caption{Curvatures and reference K\"ahler volumes for the representatives
of Table~\ref{examples-all-chambers}, in the same row order. A unit volume
in a zero-curvature slot is a choice of torus volume.}
\label{np-normalization-data-table}
\end{table}

For each row, choose the temporary scale $\fluxscale_0=|N_1|$ and form
$p_i=N_i/(\fluxscale_0\nu_i)$ and
$g_i=M_i/(2\fluxscale_0\nu_j\nu_\ell)$. Apply
\eqref{np-inverse-matrix}--\eqref{np-inverse-parameters}, denoting the
resulting interval length by $\delta_0$ and its pole ratios by
$z_i^\pm$. To reconstruct the profile on $0\leq t\leq1$, set
\begin{equation*}
 \bar c_i=\frac{c_i}{\delta_0},\qquad
 \bar d_i=\frac{d_i}{\delta_0^2},\qquad
 \bar z_i^\pm=\frac{z_i^\pm}{\delta_0},\qquad
 \bar Q_i(t)=K_it^2-4\bar c_it-4\bar d_i.
\end{equation*}
The normalized master equation uses
$\bar R=-\sum_i k_i\bar Q_i$ and $\bar P=\prod_i\bar Q_i$.
Its boundary data are
\begin{equation*}
 \bar H(0)=\bar H(1)=0,\qquad
 \bar H'(0)=4\prod_i\bar z_i^-,\qquad
 \bar H'(1)=4\prod_i\bar z_i^+.
\end{equation*}
Here $\bar H(t)=\delta_0^{-4}H(\delta_0t)$, and the physical scale in
this gauge is $\bar\fluxscale=\fluxscale_0\delta_0$. Thus, the original
radial gauges and their scale values are unnecessary for reproducing the
profiles or their scale-invariant deviations. The genus factors in the
negative-curvature rows cancel from $p_i$ and $g_i$, so one may set
$q_g=q_h=1$ when computing these radial profiles.

The regularization used in section~\ref{np-profile-deviation} is obtained
by writing $\bar H=t(1-t)Z^2$ and
$\xi=\log(t/(1-t))$. Its limiting boundary values are
$Z(-\infty)=\sqrt{\bar H'(0)}$ and
$Z(+\infty)=\sqrt{-\bar H'(1)}$, which specify the boundary
conditions for the numerical integration. 

\section{Endpoint regularity}\label{functional-analysis}
\label{np-endpoint-regularity}

The analytic implicit-function theorem gives $Y=b_q+\eta(q)$ with
$\eta\in H^2(\mathbb R)$. Consequently, $\eta,\eta_\xi\to0$ as
$\xi\to\pm\infty$. To obtain regularity in $r$, consider the left pole
and write $Y=y_-+\upsilon$, $p=\upsilon_\xi$. Equation
\eqref{np-analytic-Y-equation-general} is equivalent to
\begin{align}
 \upsilon_{\xi\xi}-\upsilon&=\mathcal F(r,\upsilon,p;q),\notag\\*
 \mathcal F&=\frac{p^2-\upsilon^2}{2(y_-+\upsilon)}-\frac12sR
                         -\frac{P-P(0)}{8(y_-+\upsilon)}.
 \label{np-endpoint-F-exact}
\end{align}
This is analytic near $(r,\upsilon,p)=(0,0,0)$ when $y_->0$.
The endpoint relation $P'(0)/4+y_-R(0)=0$ cancels its linear $r$ term,
and its other linear terms vanish. Hence
\begin{equation}
 \mathcal F=O\bigl(r^2+r(|\upsilon|+|p|)+|\upsilon|^2+|p|^2\bigr).
 \label{np-endpoint-F-estimate}
\end{equation}

Adjoining $r_\xi=r(1-r)$ gives an analytic autonomous system
\begin{equation}
 r_\xi=r(1-r),\qquad \upsilon_\xi=p,\qquad
 p_\xi=\upsilon+\mathcal F(r,\upsilon,p;q).
 \label{np-endpoint-autonomous-system}
\end{equation}
Its equilibrium at the origin has eigenvalues $1,1,-1$.
The endpoint orbit tends to that equilibrium as $\xi\to-\infty$ and
therefore lies on its analytic unstable manifold, with analytic parameter
dependence \cite{CabreFontichLlave2003II,CabreFontichLlave2005III}.
In eigencoordinates $a=(\upsilon+p)/2$ and $b=(\upsilon-p)/2$, this manifold is
$b=\Phi(r,a;q)=O((|r|+|a|)^2)$, and the restricted equation is
\begin{equation}
 a_\xi=a+\mathcal N(r,a;q),\qquad
 \mathcal N=O(r^2+r|a|+a^2).
 \label{np-endpoint-a-flow}
\end{equation}
Variation of constants on the unstable manifold shows that $a/r$ has a
finite endpoint limit. For example, hyperbolic decay first gives
$|(r,a)|=O(e^{(1-\varepsilon)\xi})$ with $0<\varepsilon<1/2$.
Then $e^{-\xi}\mathcal N$ is integrable at $-\infty$, while
$r=e^\xi+O(e^{2\xi})$.

Set $a=rA_e$. The equation becomes the nonsingular analytic ODE
\begin{equation}
 \frac{dA_e}{dr}=
 \frac{A_e+\widehat{\mathcal N}(r,A_e;q)}{1-r},\qquad
 \widehat{\mathcal N}(r,A_e;q)=\frac{\mathcal N(r,rA_e;q)}{r^2}.
 \label{np-endpoint-a-rA}
\end{equation}
Analyticity and the quadratic vanishing of $\mathcal N$ extend the quotient
through $r=0$. For every finite initial value $A_e(0)=\alpha_-$,
the resulting solution is analytic, and
$\Phi(r,rA_e;q)=r^2B_e(r,A_e;q)$ with $B_e$ analytic. Thus, \begin{equation}
 Y(r)=y_-+rA_e(r)+r^2B_e(r,A_e(r);q)
      =y_-+\alpha_-r+O(r^2),
 \label{np-endpoint-z-analytic}
\end{equation}
and there is no logarithmic endpoint term. The same argument applies at the
right pole after $r\leftrightarrow1-r$.

To connect this local statement to the global parameter dependence,
evaluate $(\eta,\eta_\xi)$ at a fixed point inside the endpoint collar.
Evaluation is continuous on $H^2$, so this state is analytic in $q$.
The analytic endpoint flow maps $(\alpha_-,q)$ to that state, with
nonzero derivative in $\alpha_-$ at the reference solution. The
finite-dimensional inverse theorem therefore makes $\alpha_-(q)$
analytic. The endpoint collars have analytic $C^k$ dependence for every
finite $k$. On a compact interior interval, the equation recovers
$Y_{\xi\xi}$ from $Y,Y_\xi$ with a positive denominator $2Y$.
Together these facts give $C^2([0,1])$ dependence of $H=sY$ and preserve
$D<0$ near the reference solution.

The local endpoint argument uses positive slopes and endpoint compatibility,
not proximity to that reference. It also applies to the barrier solution
in appendix~\ref{np-positive-certificate}, where $Y-y_-=O(r)$ and
$Y_\xi=O(r)$ supply the required limiting orbit. Near a fixed reference
solution the same proof applies with $k_2,k_3$ included among the analytic
parameters.

\section{The fully positive finite-ratio patch}
\label{np-positive-certificate}

This appendix proves the finite-ratio statement in
subsection~\ref{np-positive-flux-patch}. The purpose is deliberately
specific. We exhibit barriers that work uniformly on the box
$\mathcal B_+$ of \eqref{np-positive-flux-box}, use them to obtain a
regular solution, and then exclude polynomial solutions throughout the
same box.

\subsection{Explicit barriers and uniform bounds}
All bounds below hold uniformly on
$\mathcal B_+$ of \eqref{np-positive-flux-box}. Write
$a=(1,x,y)$, $b=(u,v,w)$, $\Sigma_f=xy+x+y$, and set
\begin{align}
 b^+&=b+(x+y,1+y,1+x),&
 \zeta&=-\frac12\mathsf L_a^{-1}b^+,\notag\\*
 \tau&=-\frac{\zeta\cdot b}{\Sigma_f},&
 z_i^\pm&=\zeta_i\pm\tau a_i+\frac{1\pm1}{4}.
 \label{np-certificate-inverse}
\end{align}
Here $\mathsf L_a$ is \eqref{np-inverse-matrix} with $p=a$.
Equation~\eqref{np-inverse-parameters} with $\delta=K_i=1$ gives
$c_i,d_i$ and exactly $I_i=\tau a_i$, $G_i^-=\tau b_i$.
Uniform rational bounds give $\tau<0$, $z_i^->0$, $z_i^+<0$ and the
three inequalities in \eqref{np-positive-Q-bounds}.

Put $s=r(1-r)$ and
$B=y_-+(y_+-y_-)r$, where
$y_-=4\prod_i z_i^-$ and $y_+=-4\prod_i z_i^+$.
A numerical profile at the centre of the box was used only to choose a
simple rational trial polynomial. The proof itself uses only the exact
function
\begin{equation}
 T(r)=\frac{13178391-4968131r-6354527r^2+6275735r^3-7049119r^4}{10^8}
 \label{np-certificate-T}
\end{equation}
and the two barriers
\begin{equation}
 H_{\rm lo}=sB+s^2(T-1/50),\qquad
 H_{\rm hi}=sB+s^2(T+1/50).
 \label{np-certificate-barriers}
\end{equation}
These barriers have the prescribed endpoint zeros and slopes. To evaluate the
residual without dividing by a vanishing endpoint coordinate, write
$\Delta=y_+-y_-$, $C_c=\sum_i c_i$ and $C_d=\sum_i d_i$. Endpoint
compatibility gives $\mathcal E_H[sB]=s^2C$, with
\begin{equation}
 C=3\Delta^2-4\Delta C_c+3y_+-C_d+4\sum_{i<j}c_ic_j-2C_c+\frac34
       +(3\Delta-C_c+\tfrac12)r+\frac14r^2.
 \label{np-certificate-C}
\end{equation}
For any polynomial $U$, direct expansion gives
\begin{align}
 \frac{\mathcal E_H[sB+s^2U]}{s^2}
 ={}&C+(R-12B)U+[6(1-2r)B-2s\Delta]U'+2sBU''\notag\\*
 &+s[-8U^2+4(1-2r)UU'+s(2UU''-(U')^2)].
 \label{np-certificate-residual-quotient}
\end{align}
The uniform bounds are
\begin{equation}
 0.050<H_{\rm lo}/s\leq H_{\rm hi}/s<0.477,\qquad
 \mathcal E_H[H_{\rm lo}]/s^2>0.039,\quad
 \mathcal E_H[H_{\rm hi}]/s^2<-0.059.
 \label{np-certificate-residual-bounds}
\end{equation}

\subsection{Existence and metric regularity}
With $H=sU^2$ and $\xi=\log(r/(1-r))$, the master equation becomes
\begin{equation}
 -U_{\xi\xi}+F(\xi,U)=0,\qquad
 F=\frac U4-\frac{sR}{4U}-\frac{P}{16U^3}.
 \label{np-certificate-semilinear}
\end{equation}
Its left-hand side is $-\mathcal E_H[H]/(4U^3)$, so
$\sqrt{H_{\rm lo}/s}$ and $\sqrt{H_{\rm hi}/s}$ are ordered sub and
supersolutions. The standard finite-interval construction
\cite{Sattinger:1972,Amann:1976}, followed by exhaustion in $\xi$, gives a
solution satisfying
\begin{equation}
 H_{\rm lo}\leq H\leq H_{\rm hi},\qquad
 |H-(sB+s^2T)|\leq\frac{s^2}{50}\leq\frac1{800}.
 \label{np-certificate-H-error}
\end{equation}
Moreover, throughout this tube,
\begin{equation}
 0.62<F_U=\frac14+\frac{sR}{4U^2}+\frac{3P}{16U^4}<3.13.
 \label{np-certificate-monotonicity}
\end{equation}
The maximum principle therefore gives uniqueness within the tube. The
endpoint expansions implied by the barriers satisfy the hypotheses of
appendix~\ref{np-endpoint-regularity}, so the solution extends smoothly
through both poles.

\paragraph{The sign of $D$.}\label{np-sheet-rigidity}
It remains to prove $D<0$. On a positive-$Q_i$ interval set
\begin{equation}
 W=\sqrt P,\qquad E_\sigma=P'+2\sigma WR,\qquad
 Z_i=W\frac{Q_i'}{Q_i}-\sigma(k_jQ_j+k_\ell Q_\ell),\qquad \sigma=\pm1.
 \label{np-boundary-E}
\end{equation}
A direct calculation using $Q_i''=2K_i$ gives
\begin{equation}
 P''-2R^2+2\sigma WR'
 =-\sum_i Z_i^2+\frac{P'}P E_\sigma,
 \qquad E_\sigma=0\ \Longrightarrow\ E_\sigma'=-\sum_iZ_i^2\leq0.
 \label{np-sheet-squares}
\end{equation}
At a zero of $E_\sigma$, equality in
\eqref{np-sheet-squares} would give $Z_i=0$ for every $i$. Put
$a_i=W(r_0)/Q_i(r_0)>0$. Then $Q_i(r_0)=a_ja_\ell$, and $Z_i(r_0)=0$
fixes $Q_i'(r_0)=\sigma(k_ja_\ell+k_\ell a_j)$. As
$Q_i''=2k_jk_\ell$, these data force
\begin{equation}
 Q_i(r_0+z)=(a_j+\sigma k_jz)(a_\ell+\sigma k_\ell z),
 \qquad W'=-\sigma R,
 \label{np-sheet-factor-Q}
\end{equation}
on the positive-$Q_i$ interval. In the fully positive sector, this
cannot satisfy the opposite pole condition $W'=\sigma R$, because
$R=-\sum_iQ_i<0$. Thus, every zero of $E_\sigma$ has strictly
negative derivative. The endpoint conditions give $E_+(0)=0$ and
$E_-(1)=0$, while $R<0$ gives $E_+(1)<0$ and $E_-(0)>0$. These
zero-crossing signs imply $E_+<0$ on $(0,1]$ and $E_->0$ on $[0,1)$.

Now set $g=2H'/W$, whose endpoint values are $1$ and $-1$, while the master
equation gives
\begin{equation}
 g=1\ \Longrightarrow\ g'=-\frac{E_+}{2P}>0,
 \qquad
 g=-1\ \Longrightarrow\ g'=\frac{E_-}{2P}>0.
 \label{np-certificate-g-crossing}
\end{equation}
Every interior crossing of either level $g=\pm1$ would therefore
have positive derivative. Since $g(0)=1$ and $g(1)=-1$, an excursion
outside $(-1,1)$ would require a crossing of one of these levels with
nonpositive derivative. Hence $|g|<1$ in the interior, and the master
equation gives $D=((H')^2-P/4)/H<0$. The same uniform bounds give
\begin{equation}
 D(0)<-7.77,\qquad D(1)<-1.82.
 \label{np-certificate-D-endpoints}
\end{equation}
Together with \eqref{np-positive-Q-bounds} and
\eqref{np-certificate-residual-bounds}, this proves regularity throughout
$\mathcal B_+$.

\subsection{Polynomial exclusion and finite verification}
For $k_i=1$, $P$ has degree six, and the leading terms of
\eqref{np-general-master} force every polynomial solution to have degree
four. An endpoint-matching quartic is therefore
\begin{equation}
 H_C=sB+C_0s^2.
 \label{np-certificate-quartic}
\end{equation}
Substitution into \eqref{np-certificate-residual-quotient} gives
\begin{equation}
 \frac{\mathcal E_H[H_C]}{s^2}
 =A_0(C_0)+A_1(C_0)r+A_2(C_0)r^2,
 \label{np-certificate-quartic-residual}
\end{equation}
where
\begin{align}
 A_2(C_0)&=\frac{(4C_0-1)(8C_0-1)}4,\notag\\
 A_1(C_0)&=3\Delta-C_c+\frac12
 +(4C_c-12\Delta)C_0-8C_0^2,\notag\\
 A_0(C_0)&=3\Delta^2-4\Delta C_c+3y_+-C_d
 +4\sum_{i<j}c_ic_j-2C_c+\frac34
 +(4C_d-12y_-)C_0.
 \label{np-certificate-quartic-coefficients}
\end{align}
The equation $A_2(C_0)=0$ restricts a possible quartic to
$C_0=1/8$ or $C_0=1/4$. Although $A_1(1/4)=0$ identically,
the remaining coefficients exclude this possibility. Uniformly on the
full box $\mathcal B_+$ of \eqref{np-positive-flux-box},
\begin{equation}
 A_1\!\left(\frac18\right)<-0.786,
 \qquad
 A_0\!\left(\frac14\right)<-1.025.
 \label{np-certificate-quartic-bounds}
\end{equation}
Neither candidate can therefore solve the master equation, so every
solution in $\mathcal B_+$ is non-polynomial independently of the tube
uniqueness argument.

These quoted uniform inequalities were checked by finite outward-rounded
rational interval arithmetic. The radial interval was
split into 1024 closed pieces, and the two residual signs were bounded by a
centred first-order enclosure in the four flux variables, while the
remaining quantities used a $3^4$ subdivision of the flux box. Representative
margins are
\begin{equation}
 \frac{\mathcal E_H[H_{\rm lo}]}{s^2}>0.0395,\qquad
 \frac{\mathcal E_H[H_{\rm hi}]}{s^2}<-0.0596,\qquad
 Q_2>0.0153,\qquad F_U>0.628.
 \label{np-certificate-verifier-margins}
\end{equation}
These bounds use the rational endpoint inverse and the explicit barriers
above.

\section{Exact flux differentiation}\label{np-exact-flux-jacobian}

For reference, this appendix records the algebraic differentiation
underlying the flux-coordinate Jacobian in \eqref{np-analytic-flux-J}.
Notice that here no interior profile is differentiated.  In the $(1,0,0)$ sector, let
$p=(A_1,B_1,A_2,B_2,A_3,B_3)$ and $q=(A_2,B_2,A_3,B_3)$.  Introduce
\begin{equation}
 W_-=\sqrt{A_1A_2A_3},\qquad
 W_+=\sqrt{(A_1+B_1)(A_2+B_2)(A_3+B_3)},
\end{equation}
so that the endpoint constraints are explicitly
\begin{align}
 F_-&=\frac{W_-}{2}\sum_i\frac{B_i}{A_i}-A_1,\notag\\*
 F_+&=\frac{W_+}{2}\sum_i\frac{B_i}{A_i+B_i}+(A_1+B_1).
 \label{np-flux-J-endpoint-explicit}
\end{align}
At $p_0=(1/8,0,1,1,2,-1)$, their $q$ derivative is
\begin{equation}
 D_q(F_-,F_+)\big|_0=
 \begin{pmatrix}-\tfrac3{16}&\tfrac14&\tfrac3{32}&\tfrac18\\
 -\tfrac3{32}&\tfrac1{32}&\tfrac3{16}&\tfrac7{16}\end{pmatrix}.
 \label{np-flux-J-endpoint-q}
\end{equation}
Multiplying by minus the inverse of \eqref{np-analytic-endpoint-J}
gives \eqref{np-flux-J-dependent}. In this sector the endpoint periods
simplify to
\begin{align}
 I_i&=-\frac14\left(\frac{W_-}{A_i}
                   +\frac{W_+}{A_i+B_i}+k_i\right),\notag\\*
 G_1^-&=-\frac{B_1}4,\qquad
 G_2^-=\frac{W_-}{4A_3}-\frac{B_2}4,\qquad
 G_3^- =\frac{W_-}{4A_2}-\frac{B_3}4.
 \label{np-flux-J-periods}
\end{align}
Differentiating the four ratios first in the six independent variables gives
\begin{equation}
 D_p\mathcal F\big|_0=
 \begin{pmatrix}
 \tfrac{17}{27}&\tfrac7{27}&-\tfrac1{16}&-\tfrac7{432}&0&-\tfrac1{216}\\
 \tfrac{17}{27}&\tfrac{10}{27}&0&\tfrac1{216}&-\tfrac1{16}&-\tfrac5{108}\\
 0&\tfrac19&0&0&0&0\\
 \tfrac5{27}&\tfrac4{27}&-\tfrac1{24}&\tfrac{11}{108}&-\tfrac1{48}&-\tfrac1{54}
 \end{pmatrix}.
 \label{np-flux-J-six-variable}
\end{equation}
The endpoint-compatible derivative is the product of this matrix with
\begin{equation}
 D_qp=\begin{pmatrix}D_q(A_1,B_1)\\ \mathbf 1_4\end{pmatrix},
 \label{np-flux-J-embedding}
\end{equation}
which gives exactly \eqref{np-analytic-flux-J}. 

The same endpoint equations determine the deformation used to exclude
polynomiality. For $q=(1,1+\epsilon,2,-1)$, they give
\begin{align}
 A_1&=\frac18+\frac14\epsilon+\frac14\epsilon^2
       +\frac{13}{144}\epsilon^3-\frac1{108}\epsilon^4+O(\epsilon^5),
 \notag\\*
 B_1&=-\frac1{16}\epsilon-\frac3{32}\epsilon^2
       -\frac{41}{576}\epsilon^3-\frac5{3456}\epsilon^4+O(\epsilon^5).
 \label{np-analytic-A1B1-expansion}
\end{align}
Substituting into $\beta=(W_+-W_-)/2$ gives
$\beta=\epsilon^3/144+\epsilon^4/864+O(\epsilon^5)$.
Thus, the first nonzero term of $\beta(B_1+3\beta)$ is
$-\epsilon^4/2304$, as used in the main proof.

\section{Finite degenerations and the flat limit}\label{finite-degenerations}

We now supply the proof behind the finite-limit classification in
section~\ref{np-finite-boundary-classification}.  The argument separates
cross-sheet collapse from positive-length limits.  The endpoint-sheet
identity and its equality case, established in appendix~\ref{np-sheet-rigidity},
are the main algebraic input, and the compactness assumptions are exactly those
stated in the main text, so they will not be repeated here.

\subsection{Cross-sheet collapse}
\label{np-cross-sheet-collision}

The left and right endpoints lie on the sheets
$E_+=P'+2\sqrt P\,R=0$ and $E_-=P'-2\sqrt P\,R=0$.  If they meet at
$r_{\rm c}$ with $P_{\rm c}>0$, both equations hold there, immediately
giving \eqref{np-boundary-collapse-condition}.  If the two sheet roots are
simple, the implicit-function theorem makes their separation linear in a
transverse continuation parameter. A multiple sheet root instead falls into
the factorized equality case \eqref{np-sheet-factor-Q}.

The leading collapse profile is equally simple.  From
\begin{equation}
 \mathcal S_H=P-4(H')^2=-4HD>0
 \label{np-boundary-S}
\end{equation}
one has $\|H\|_\infty=O(\Delta_r)$ when
$\Delta_r=r_+-r_-\to0$ and $P\to P_{\rm c}>0$.  Set
$r=r_-+\Delta_r t$ and $H=\Delta_r h_{\Delta_r}(t)$.  Assuming
$h_{\Delta_r}\to h_0$ in $C^2([0,1])$, the limiting master equation is
$-(h_0')^2+2h_0h_0''+P_{\rm c}/4=0$ with endpoint slopes
$\pm\sqrt{P_{\rm c}}/2$.  Hence
\begin{equation}
 h_0(t)=\frac{\sqrt{P_{\rm c}}}{2}t(1-t),\qquad
 H_{\max}=\frac{\sqrt{P_{\rm c}}}{8}\Delta_r+o(\Delta_r).
 \label{np-boundary-collapse-profile}
\end{equation}
In particular, $D\sim-2\sqrt{P_{\rm c}}/\Delta_r$ and
$\omega^6\sim\Delta_r^2$.  The fibre tends to a round two-sphere in the
unwarped bracket, while the full warped metric degenerates.  The finite
limits \eqref{np-boundary-collapse-observables} follow directly from the
endpoint formulae and do not require the extra rescaled $C^2$ convergence.

\subsection{A zero of \texorpdfstring{$Q_i$}{Qi}}
\label{np-Q-pinch}

Suppose $Q_i$ has positive endpoint values and develops an isolated interior
zero while $Q_j,Q_\ell$ stay positive.  Write
$Q_j^*=Q_j(r_i^*)$ and $Q_\ell^*=Q_\ell(r_i^*)$.  The quadratic positivity criterion
\eqref{np-inverse-Q-test} forces a double zero and $K_i=k_jk_\ell>0$.
Writing $\rho=r-r_i^*$ gives
$P=p_2\rho^2+o(\rho^2)$, with $p_2=K_iQ_j^*Q_\ell^*>0$.
If $H(r_i^*)>0$, the master equation and $\mathcal S_H\geq0$ imply
$D=H'=0$ there, but then
$\mathcal S_H=(p_2-R_*^2)\rho^2+o(\rho^2)$, whereas
\begin{equation}
 R_*^2=(k_jQ_j^*+k_\ell Q_\ell^*)^2
 \geq4k_jk_\ell Q_j^*Q_\ell^*=4p_2,
 \label{np-Qpinch-AMGM}
\end{equation}
which is impossible.  Thus, $H=H'=0$ at the pinch, and, inserting
$H=\alpha\rho^2+o(\rho^2)$ into the master equation, gives $R_*<0$, or, equivalently,
$k_j,k_\ell>0$.  This proves
\begin{equation}
 Q_i=Q_i'=H=H'=0,\qquad k_j,k_\ell>0.
 \label{np-Qpinch-positive-pair}
\end{equation}
Moreover, $H\sim\rho^2$ and $Q_i\sim\rho^2$, while $D$ tends to a finite negative
limit.  In the physical metric, the radial and $C_i$ coefficients grow as
$|\rho|^{-4/3}$ and the remaining coefficients collapse as $|\rho|^{2/3}$, so
the pinch lies at finite proper distance and is singular.
A proper-distance coordinate $t$ on either side satisfies
$t\propto|\rho|^{1/3}=|r-r_i^*|^{1/3}$. The asymptotic metric therefore
has the schematic behaviour
$ds^2\sim dt^2+t^{-4}ds^2(C_i)+t^2ds^2_{\perp}$.
The unequal warp exponents force curvature to diverge as $t\to0$.  The degeneration
therefore mixes decompactification of $C_i$ with collapse of the
complementary directions. Its local scaling is incompatible with a smooth bolt
or orbifold quotient and does not resemble a regular brane throat within this
ansatz.

\subsection{Excluding an independent \texorpdfstring{$D$}{D} boundary}
\label{np-D-exclusion}

Assume that the limiting interval has positive length and all $Q_i$
remain uniformly positive. The limiting profile is nonnegative. An
interior zero would have $H'=0$ and would contradict the master equation
because $P>0$. Thus, $H>0$ throughout the open interval, and the assumed
smooth interior convergence permits differentiation there.

For $\sigma=\pm1$, define
\begin{equation*}
 \Phi_\sigma=P''-2R^2+2\sigma\sqrt P\,R'.
\end{equation*}
Differentiating the master equation once and twice gives
\begin{align*}
 HD'+H'R+\tfrac14P'&=0,\\
 HD''+H'D'+H''R+H'R'+\tfrac14P''&=0.
\end{align*}
If $D$ first reaches zero at an interior point $r_0$, then
$D=D'=0$ and $D''\leq0$ there. The master equation gives
$H'=\sigma\sqrt P/2$ and $H''=-R/2$. The first differentiated equation
then gives $E_\sigma=0$, while the second gives
\begin{equation*}
 \Phi_\sigma(r_0)=-4H(r_0)D''(r_0)\geq0.
\end{equation*}
On the other hand, \eqref{np-sheet-squares} gives
$\Phi_\sigma(r_0)=-\sum_iZ_i(r_0)^2\leq0$.
Hence every $Z_i(r_0)$ vanishes.

Pointwise equality already determines the quadratic data. Set
\begin{equation*}
 a_i=\sqrt{\frac{Q_j(r_0)Q_\ell(r_0)}{Q_i(r_0)}}>0,
 \qquad \{i,j,\ell\}=\{1,2,3\},
\end{equation*}
then $Q_i(r_0)=a_ja_\ell$, and $Z_i(r_0)=0$ fixes
$Q_i'(r_0)=\sigma(k_ja_\ell+k_\ell a_j)$. Since
$Q_i''=2k_jk_\ell$, these values and derivatives determine the entire
quadratic and give exactly \eqref{np-sheet-factor-Q}. On the connected
positive-$Q_i$ interval, the factors cannot cross zero, so
$W'=-\sigma R$ throughout. The profile defined by
$H_*'=\sigma W/2$ and $H_*(r_0)=H(r_0)$ consequently has $D_*=0$
and solves the master equation. At $H>0$, the equation is a regular
second-order ODE, so that uniqueness for the initial data $(H,H')$
identifies the limiting profile with $H_*$ throughout the interior.
As it is strictly monotone, it cannot have two distinct endpoint zeros.

At a pole $r_e$, the nonzero slope and endpoint compatibility give the
smooth germ of appendix~\ref{np-endpoint-regularity}. If $D(r_e)=0$,
then $H''(r_e)=-R(r_e)/2$ and the differentiated equation above yields
\begin{equation*}
 D'(r_e)=-\frac{\Phi_\sigma(r_e)}{4H'(r_e)},
 \qquad H'(r_e)=\frac{\sigma\sqrt{P(r_e)}}2.
\end{equation*}
The pole condition gives $E_\sigma(r_e)=0$, hence
$\Phi_\sigma(r_e)=-\sum_iZ_i(r_e)^2\leq0$. At the left pole,
$H'>0$ implies $D'\geq0$, whereas the one-sided condition $D\leq0$
requires $D'\leq0$. At the right pole the two inequalities reverse.
Thus, $\Phi_\sigma(r_e)=0$ at either pole, and the same pointwise
argument fixes the factorized quadratics. The smooth endpoint germs in
appendix~\ref{np-endpoint-regularity} are uniquely determined by the
endpoint data and $H''(r_e)$, so that the limiting profile agrees
locally with the solution $H_*'=\sigma W/2$, $H_*(r_e)=0$, and interior
ODE uniqueness propagates this agreement. Monotonicity again excludes
the opposite pole. Consequently, $D<0$ cannot fail independently while
all $Q_i$ stay positive.

\subsection{Completion of the finite-limit argument}
\label{np-finite-boundary}

If the interval does not collapse and no $Q_i$ vanishes, the limiting
quadratics have a uniform positive margin.  The limiting profile cannot
acquire an interior zero, since $H=H'=0$ would force $P=0$ by the master
equation.  The preceding subsection gives $D<0$, and
Appendix~\ref{np-endpoint-regularity} supplies smooth endpoint germs.
The limiting solution is therefore regular, which proves the two
alternatives stated in subsection~\ref{phase-boundaries}.  Endpoint or
simultaneous zeros of the $Q_i$ belong to the second alternative, but are not
further classified here.  In the non-flat nonnegative-curvature sectors, $R<0$ also excludes
a pure collapse with $P_{\rm c}>0$.

\subsection{The smooth flat stratum}\label{np-flat-corner}

The completely flat family is smooth rather than singular.  Notice that, if we put
$k_i=\epsilon\kappa_i$ and
$Q_i=Q_i^{(0)}+\epsilon(A_ir+B_i)+O(\epsilon^2)$, with $Q_i^{(0)}>0$, in the gauge
$r_-=0$, $r_+=1$,  the two linearized endpoint conditions reduce to
\begin{equation}
 \sum_i\kappa_iQ_i^{(0)}=0,
 \qquad
 \sum_i\frac{A_i}{Q_i^{(0)}}=0.
 \label{np-flat-tangent-conditions}
\end{equation}
Thus, every nonzero finite-width curvature tangent contains both signs.  The
linearized change of $H$ is only an overall rescaling, so shape changes begin
at second order.  At the flat point
\begin{equation}
 I_i=-\frac{\sqrt{Q_1^{(0)}Q_2^{(0)}Q_3^{(0)}}}{2Q_i^{(0)}},\qquad G_i^-=0,
 \qquad
 (x,y,u,v)=\left(\frac{Q_1^{(0)}}{Q_2^{(0)}},\frac{Q_1^{(0)}}{Q_3^{(0)}},0,0\right).
 \label{np-flat-flux-locus}
\end{equation}
The flat family is therefore a two-dimensional stratum of the four-ratio
space.  Its curvature-tangent condition has the same algebraic form as
\eqref{np-collapse-flux-wall}, but the geometry is entirely different. The
flat branch retains a finite smooth interval whereas the collapse wall has
coincident endpoints.

\section{Numerical continuation data}\label{continuation-data}

The paths in section~\ref{nonpoly-continuation} use the following rounded
reduced constants. Exchanging factors two and three in the last row gives
the $(1,-1,1)$ representative obtained by permuting the last two factors
of the $(+,+,-)$ row in Table~\ref{examples-all-chambers}.
\begin{center}\footnotesize
\begin{tabular}{@{}c l@{}}
\toprule
$(k_1,k_2,k_3)$ & $(c_1,c_2,c_3)$ and $(d_1,d_2,d_3)$\\
\midrule
$(1,-1,-1)$ & $(1.17,0.653013635971737,1.12)$\\
 & $(-1.19003310423892,-1.45396981128029,1.06436142040867)$\\
$(1,-1,0)$ & $(0.9,0.0987714513021423,0)$\\
 & $(-1.09915868537647,-0.997952816304149,-1.00068136487580)$\\
$(1,1,-1)$ & $(0.7,-1.18145589020189,1.6)$\\
 & $(-1.11993903785466,-1.87860841172288,-1.17892148758334)$\\
\bottomrule
\end{tabular}
\end{center}

Endpoint roots are tracked using $P'^2-4PR^2=0$, retaining the
physical slopes $H'_- =+\sqrt{P_-}/2$ and
$H'_+=-\sqrt{P_+}/2$. The two values $H''_\pm$ are the shooting
parameters. At the computed continuation nodes the inward solutions are
matched, and a radial mesh is used to check $H>0$, $Q_i>0$ and $D<0$.
These checks are numerical and do not bound the intervals between nodes.

The first segment fixes the $(1,-1,-1)$ constants and varies
$\kappa=k_3$ from $-1$ to zero. At zero, the inward integrations agree
to a few parts in $10^{-12}$. A radial shift sets $c_3=0$, after which
interpolation of the remaining constants to the rational $(1,-1,0)$
representative is solved at 51 intermediate points. The final segment
sets $\kappa=\vartheta$, with $\vartheta\in[0,1]$ also parametrizing
the interpolation of the remaining constants from that rational point to the permuted $(1,-1,1)$ representative from Table~\ref{examples-all-chambers}, using 81 continuation nodes.
The sampled profiles pass the matching and positivity checks.
Representative values for the first and final segments are
\begin{center}\small
\begin{tabular}{@{}c r r r r r@{}}
\toprule
Segment & $\kappa$ & $r_-$ & $r_+$ & $H_{\max}$ & $\min\omega^6$\\
\midrule
Fixed constants & $-1$ & $-3.01822040$ & $-1.77535146$ & $1.01319$ & $0.1426$\\
 & $-0.75$ & $-2.94502229$ & $-1.85908217$ & $1.08290$ & $0.2770$\\
 & $-0.50$ & $-2.85432674$ & $-1.95901753$ & $1.05053$ & $0.3474$\\
 & $-0.25$ & $-2.74554573$ & $-2.07662366$ & $0.88753$ & $0.2948$\\
 & $0$ & $-2.62104019$ & $-2.21154085$ & $0.58984$ & $0.1448$\\
\midrule
Interpolated & $0$ & $-1.467707$ & $-0.501357$ & $1.14208$ & $0.66423$\\
 & $0.25$ & $-0.966218$ & $-0.548264$ & $0.59528$ & $0.16069$\\
 & $0.50$ & $-0.747866$ & $-0.506977$ & $0.38069$ & $0.05580$\\
 & $1$ & $-0.582713$ & $-0.247070$ & $0.61376$ & $0.09980$\\
\bottomrule
\end{tabular}
\end{center}
For the interpolated segment the matching residuals are of order
$10^{-12}$--$10^{-13}$, and no endpoint or shooting degeneracy is seen at
the sampled nodes.  Notice that these are merely numerical diagnostics rather than validated
continuation bounds.

If the initial $(1,-1,-1)$ constants are instead held fixed while
$\kappa$ increases beyond zero, the endpoint roots merge near
\begin{equation}
 \kappa_{\rm c}=0.358514139306698\ldots,\qquad
 r_{\rm c}=-2.426750066\ldots .
 \label{np-cont-collapse}
\end{equation}
The computed data have $P'_{\rm c}\simeq R_{\rm c}\simeq0$,
$P_{\rm c}\simeq144.9177046>0$ and
\begin{equation}
 (\partial_rE_+)_{\rm c}\simeq-495.2821430,\qquad
 (\partial_rE_-)_{\rm c}\simeq-420.8353376.
 \label{np-cont-collapse-diagnostic}
\end{equation}
The limiting flux ratios are $x\simeq0.9817732775$ and
$y\simeq19.3111845295$. A second reported collapse has critical
curvatures $(0.1601847268\ldots,-1,-1)$ and limiting ratios
$x\simeq20.0749254011$, $y\simeq9.0603229928$. These two datasets
are consistent with the necessary wall
\eqref{np-collapse-flux-wall} at the printed precision. They concern
cross-sheet collapse and give no realization of the isolated interior
$Q_i$ pinches analysed in appendix~\ref{np-Q-pinch}.

\end{document}